\documentclass[a4paper]{article}

\usepackage[pages=all, color=black, position={current page.south}, placement=bottom, scale=1, opacity=1, vshift=5mm]{background}

\usepackage[margin=1in]{geometry} 

\usepackage{amsmath}
\usepackage{amsthm}
\usepackage{amssymb}
\usepackage{lineno}
\usepackage{url}

\usepackage[utf8]{inputenc}

\usepackage [autostyle, english = british]{csquotes}
\MakeOuterQuote{"}

\usepackage{graphicx, color}

\usepackage{lineno}

\title{The Rising Entropy of English in the Attention Economy}

\author{Charlie Pilgrim$^{1, 2, 3, *}$ \and Weisi Guo$^{3,4}$ \and Thomas T. Hills$^{3,5}$}

\date{
	$^1$The Mathematics of Real-World Systems CDT, The University of Warwick, Coventry, CV4 7AL, UK\\
    $^2$Experimental Psychology, University College London, London, WC1H 0DS, UK\\ 
	$^3$The Alan Turing Institute, London, NW1 2DB, UK\\ 
    $^4$Human Machine Intelligence Group, Cranfield University, Bedford, MK43 0AL, UK\\
	$^5$Department of Psychology, The University of Warwick, Coventry, CV4 7AL, UK\\
    $^*$Corresponding author: Charlie Pilgrim, c.pilgrim@ucl.ac.uk\\ 
}

\begin{document}
\maketitle

\begin{abstract}
    We present evidence that the word entropy of American English has been rising steadily since around 1900, contrary to predictions from existing sociolinguistic theories. We also find differences in word entropy between media categories, with short-form media such as news and magazines having higher entropy than long-form media, and social media feeds having higher entropy still. To explain these results we develop an ecological model of the attention economy that combines ideas from Zipf's law and information foraging. In this model, media consumers maximize information utility rate taking into account the costs of information search, while media producers adapt to technologies that reduce search costs, driving them to generate higher entropy content in increasingly shorter formats. 
\end{abstract}

\linebreak

\noindent

\section*{Introduction}


Word entropy is a measure of the amount of repetition (low entropy) or novelty (high entropy) in word distributions. Empirical word distributions typically follow Zipf's law, which describes a power law between a word's observed frequency and that word's rank in the frequency distribution \cite{zipf1949human}. This empirical power law is remarkably stable with an exponent around 1 
\cite{Bentz2015Jun, baixeries2013evolution, Ferreri.Cancho2005Mar}. The stability of Zipf's law suggests some underlying mechanism, and Zipf himself hypothesised a \emph{principle of least effort} between speakers and listeners. More recently this principle has been expanded to show that power laws in word distributions can emerge from a balance between maximising the benefits of receiving highly informative messages (preferred by listeners) and minimising the costs of generating high word entropy text (preferred by speakers) \cite{Ferreri.Cancho2005Mar}. 

In recent times this balance between the efforts of listeners and speakers has changed. Modern communication systems have transformed the way that we share and consume information, in particular by increasing the accessibility of information \cite{hills2019dark}. In the words of Herbert Simon this creates a "poverty of attention" \cite{simon1955behavioral}, such that media producers must compete for the limited resource of human attention  \cite{evans2020economics, ciampaglia2015production, terranova2012attention}. This dynamic has been called \textit{the attention economy}, a combination of forces influencing the production and consumption of information, with consequences including a shortening collective attention span \cite{Lorenz-Spreen2019Apr}. If information adapts to the balance between the preferences of media producers and consumers, then increased competition for attention tips the balance toward the preferences of the consumers. That is, information markets (the distribution of available content) should rise in information density, and specifically, entropy.  

We can envision this adaptive process in terms of \emph{information foraging} \cite{pirolli1999information, sandstrom1994optimal}. Information foraging  describes how people search for and consume information in different environments, including web browsing \cite{pirolli2009information} software debugging \cite{lawrance2010programmers, lawrance2010reactive, piorkowski2013whats}, and the design of information and social environments \cite{pirolli2009information, piorkowski2013whats, bhowmik2015optimal}. The basic rationale of this approach is borrowed from ecological models of foraging, which have been shown to be appropriate to a wide range of search problems ranging from spatial foraging to cultural evolution \cite{hills2015exploration}. Indeed, handling the exploration versus exploitation trade-off that is common to all of these environments has been proposed to be a defining selective force in the evolution of cognition \cite{hills2006animal,todd2020foraging}. 


In what follows, we first investigate the evolution of information across a wide variety of media sources over the last two centuries, a time marked by increasing media competition. We show how this reveals a characteristic pattern of rising entropy that affects different categories of media in different ways (e.g., books versus news versus social media). We then create a model of the attention economy that  expands on existing models of information foraging to incorporate competition for human attention between media producers. This model explains both the general increase in word entropy and the differences in word entropy across categories.

\section*{Materials and Methods}

\textbf{Text Corpora}

To investigate the recent history of information evolution we examine a variety of text corpora. The Corpus of Historical American English (COHA) \cite{Davies2012Nov} has $116,614$ texts spanning the 1810s to 2000s, balanced between categories of fiction ($n=11,010$), non-fiction ($n=2,635$), news ($n=41,677$) and magazines ($n=61,292$). The Corpus of Contemporary American English (COCA) has over 150,000 texts from between 1990 to 2008 split between fiction, popular magazines, newspapers, academic journals and spoken word \cite{davies2009385+}. For our analysis we used a publicly available sample of COCA with $2,362$ texts split between categories of fiction ($n=275$), academic journals ($n=266$), news ($n=872$) and magazines ($n=949$). The British National Corpus (BNC) contains $8,098$ texts from between 1960 and 1993 including written categories of fiction ($n=904$), academic prose ($n=994$), newspapers ($n=972$), non-academic prose and biography, other published materials and unpublished materials \cite{ByLouBurnard2007Jan}. Fiction and newspapers are common categories across the corpora. Magazines are a common category between COHA and COCA. We grouped as non-fiction the categories of COHA non-fiction, COCA academic journals and BNC academic prose. 

The text sample data was cleaned before analysis in a standard way \cite{gerlach2020standardized}. COHA and COCA are similar formats and so followed the same procedure. For both:

\begin{itemize}
    \item Stripped any headers not a part of the main text samples.
    \item Removed any XML text tags.
    \item Removed any sentences that contained "@" symbols. COHA and COCA randomly replace words with @ symbol in groups of ten for copyright reasons \cite{rudnicka2018variation}. 
    \item Removed apostrophes and extra whitespace.
    \item Used python's natural language toolkit (nltk) package to convert text to tokens \cite{bird2009natural}.
    \item Selected the last 2000 tokens (words) of the text sample for processing. This avoids, as much as possible, anomalous text that sometimes appears at the start of text samples such as a contents section.
\end{itemize}

For the BNC data, python's natural language toolkit package comes with a BNC corpus reader \cite{bird2009natural}, which was used to extract tokens. The only other treatment was to remove extra whitespace and apostrophes as with COCA and COHA. 

The cleaned datasets had the following surviving sample counts with $N \geq 2000$ words:

\begin{itemize}
    \item COHA total $n=22,253$. Fiction $n=8,164$, non-fiction $n=2,046$, news $n=725$, magazines $n=11,318$.
    \item COCA total $n=985$. Fiction $n=167$, non-fiction $n=166$, news $n=39$, magazines $n=133$. 
    \item BNC total $n=1,319$. Fiction $n=447$, non-fiction $n=477$, news $n=395$. 
\end{itemize}

The COHA dataset was analysed as a timeseries, so requires a large number of samples. The BNC and COCA, being corpora from much narrower time ranges,  were analysed as distributions and as such require less samples. 

\textbf{Social Media Data}

We also investigated social media. The Twitter dataset consisted of 1.6 million tweets scraped from the twitter API between April and June 2009 \cite{go2009twitter} and available online at \url{https://www.kaggle.com/kazanova/sentiment140}. To simulate a Twitter feed the tweets were chronologically collated to create $n=1000$ text samples with $N \geq 2000$ words each. 

For Reddit, we aimed to capture text samples that were representative of the text a user would see when visiting the site. To achieve this we used Reddit's API to download posts from the Reddit homepage feed at \url{https://oauth.reddit.com/.json}. Following Reddit's API rules, we first registered an app and all requests were authenticated with OAuth2. We downloaded 10,000 posts in JSON format in this way. We extracted the text from the posts and combined them to create $n=90$ text samples with length $N \geq 2000$ words each. During processing we found a small number of non-English posts in the feed, which were removed. 

The social media data was then cleaned:  

\begin{itemize}
    \item Removed apostrophes and extra whitespace.
    \item Removed any urls.
    \item Removed hashtags and usernames i.e. any words containing "@" or "\#".
    \item Used python's natural language toolkit (nltk) package \cite{bird2009natural} to convert the collated samples into a list of tokens, and the last 2000 tokens taken. 
\end{itemize}

Social media statuses are by nature short and are usually much smaller than $N  = 2000$ words, and lexical measures of short text samples have little meaning. Our analysis is on the level of the social media feed and we generated large text samples through the collation of posts. This kind of collation will naturally create text samples with high lexical diversity. This isn't a flawed analysis --- the high information density of a social media feed is related to the collation of statuses and how people actually consume social media.  

\textbf{Measures of information evolution}

Information evolution is measured using unigram word entropy. For robustness we also analysed the type token ratio and Zipf exponent of text samples, which are also measures of lexical diversity \cite{Bentz2015Jun}. The lexical measures are all sensitive to sample size, so we used truncated text samples to $N=2000$ words. 

Empirical unigram word entropy, $H_1$, is a function of the relative frequencies of each word, $f_i$, summed over the set of $W$ unique words in the text sample. We use the maximum likelihood or plug-in estimator, which has the benefit of being simple and well known. And it has been shown to correlate well with more advanced estimators \cite{bentz2017entropy}.

\begin{equation}
    H_1 = - \sum_{i=1}^W f_i log_2 f_i \,.
\end{equation}

Type token ratio (TTR) is the number of unique words (types) divided by the total words (tokens) in a text sample. 

\begin{equation}
    TTR = \dfrac{\# types}{\# tokens} \,.
\end{equation}

Words in natural language are typically approximately distributed as a power law distribution between type frequency, $f_i$, and type rank in that frequency distribution, $r(f_i)$ \cite{clauset2009power}. This power law is parameterised by the Zipf exponent, $\alpha$, which describes the steepness of the distribution in log space. Maximum likelihood estimation was used to estimate the Zipf exponent \cite{clauset2009power}. This estimator has the benefit of being widely used and well known. It shows bias (as do all Zipf estimators \cite{pilgrim2020bias}), but the bias is systematic so can be ignored for the purpose of comparision of text samples. 

\begin{equation}
    f_i \propto r(f_i)^{- \alpha} \,.
\end{equation}

Each of the measures were applied once to the same set of distinct text samples. 

\textbf{Timeseries Breakpoint Analysis}

The Corpus of Historical American English (COHA) provides historical text samples across fiction, non-fiction, news and magazines categories. The type token ratio, word entropy and Zipf exponent were calculated for each text sample with over 2000 words.

For each media category and lexical measure, the results were binned into years and the median taken each year. The median was used to reduce the effect of outliers (similar results were found when using the mean). These were plotted on a scatterplot (see  Supplementary Information). 

Visually, the scatterplots are suggestive of some change in the gradient of the lexical measure in time. In order to estimate the location of these breakpoints, we used python's piecewise-regression package \cite{pilgrim2021piecewise} with default settings. The regression fits and locations of breakpoints are shown in the scatterplots in the Supplementary Information. 

We ran a similar analysis with the categories combined. In order to combine the categories, we first took means for each year and category and then took the mean across categories for each year. It is more natural to use means than medians when combining categories, and the influence of outliers is smaller as there is more data than in the individual categories. The scatterplot and piecewise-regression fit for the combined word entropy is shown in the Supplementary Information. 

\textbf{Timeseries Trend Analysis}

For each category and lexical measure, trend analyses were carried out on the annual median values. This was done between the years 1900 and 2009 (the last year of data). KPSS and MK tests were carried out for each measure and media category in COHA (full results in Supplementary Information). 

The Kwiatkowski–Phillips–Schmidt–Shin (KPSS) test assumes the null hypothesis of a stationary timeseries. p-values below 0.05 mean that we can reject this hypothesis at 5\% significance and provide evidence of a trend. The test was applied using python's statsmodels package \cite{seabold2010statsmodels}. 

The Mann-Kendall (MK) test is a non-parametric trend test \cite{hussain2019pymannkendall}. The test assumes no serial correlation i.e. errors in one observation do not predict errors in other observations \cite{hussain2019pymannkendall}.  The text corpora are constructed from independent text samples so this is a reasonable assumption. The null hypothesis is that the data has no trend, and the p-value tells us the probabilty that the data was observed under the null hypothesis. At 5\% significance we reject the null hypothesis if $p<0.05$. The test was carried out using python's pymannkendall package \cite{hussain2019pymannkendall}

In the Supplementary Information we calcaulte Pearson's R between magazine circulation and word entropy. 

\textbf{Timeseries Smoothing}

While we included scatterplots for annual binned data in the Supplementary Information, the trends are easier to see visually with a smoothed timeseries. For Figures  \ref{fig:timeseries} and \ref{fig:media-categories} the timeseries was smoothed using a moving average with measures of text samples from $\pm$ 5 years. The 95\% confidence interval was calculated as the standard error of this mean calculation multiplied by 1.96 (assuming normally distributed errors). For each lexical measure, the mean was plotted for each year with the confidence interval region shaded. We only included years where we had a minimum of 10 data points within the window. 

We report the smoothed timeseries for each of the COHA text categories, as well as the categories combined. The timeseries for media categories were combined by taking an average across the timeseries annual means for the media categories that had a value for that year. The 95\% confidence interval was again calculated as 1.96 times the standard error. For each year, the standard error of the estimate of the mean, $SE_{\bar{X}}$ was computed based on the delta method, 

\begin{equation}
    SE_{\bar{X}} = \dfrac{\sqrt{\sum_{i=1}^{n} SE_i^2}}{n} \,,
\end{equation}

with $n$ depending on how many media categories had values for the annual mean each year.

\textbf{Differences Between Media Categories}

We looked at the distributions of the lexical measures within media categories in COCA, the BNC and COHA (restricted to 2000-2007 to avoid the effect of historical changes). To test for differences between the groups we carried out ANOVA tests across categories within each corpora separately for each of the lexical measures. At 5\% significance, $p<0.05$ provides evidence that the media categories are drawn from different underlying population distributions. The tests were carried out using python's statsmodels package \cite{seabold2010statsmodels} 

For visualisation, the distributions of word entropy for each media category are shown as a kernel density estimate with the bandwidth determined by the Scott rule and the density trimmed to the data range.   

\textbf{US Magazine Circulation}

The data for magazine circulation numbers (reported in the Supplementary Information) were taken from Sumner's "The Magazine Century American Magazines Since 1900" \cite{sumner2010magazine} Chapter 1, which are attributed to data originally from the Audit Bureau of Circulation. This data source does not track all US magazines, but does track well-known magazines. The data was plotted without further treatment. 



\section*{Results}

\subsection*{The Rising Entropy of American English}

\begin{figure}[ht]
		\centering
		\includegraphics[width=1\textwidth]{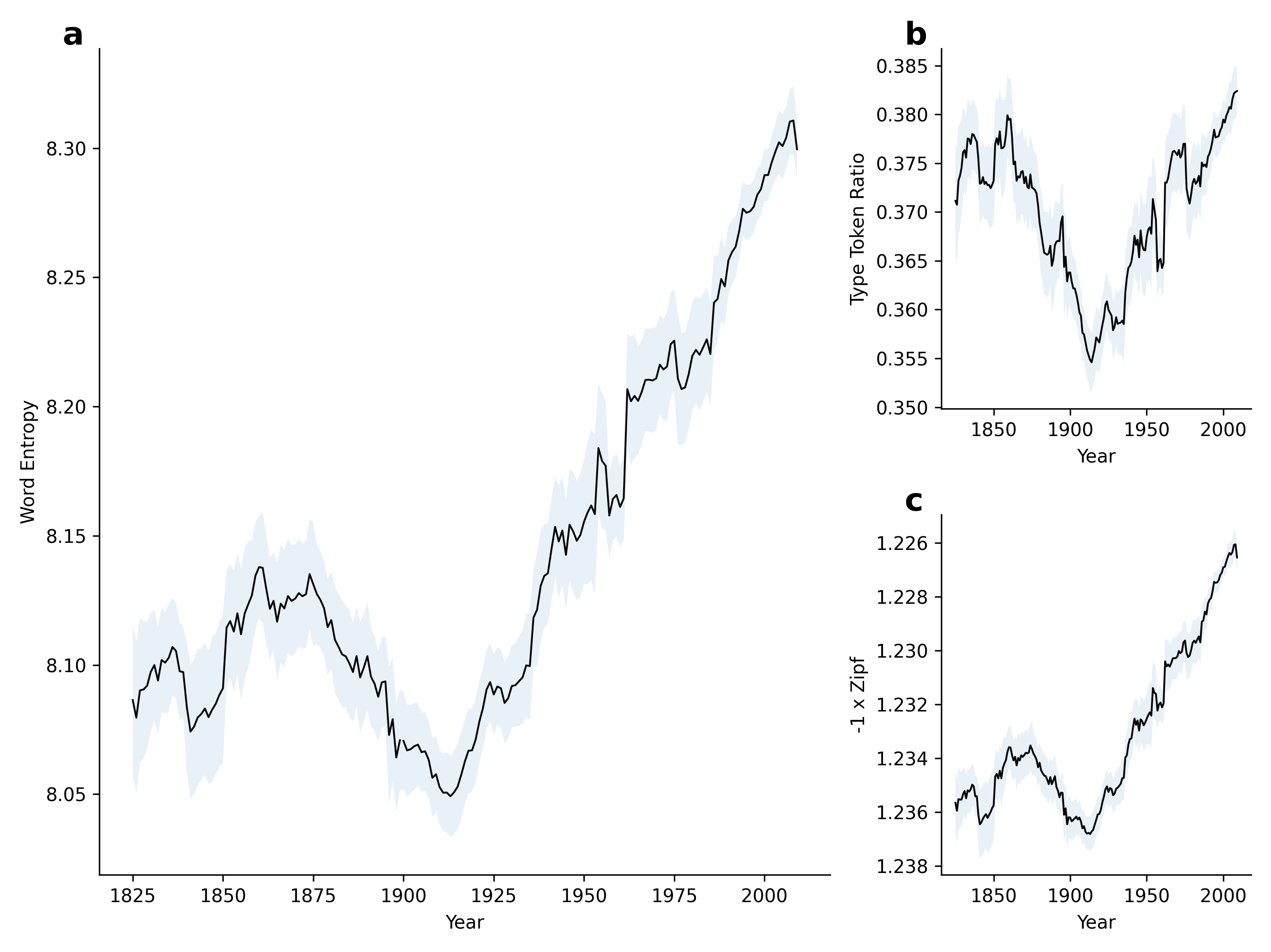}
		\caption{Lexical diversity of text samples in the Corpus of Historical American English as measured by a) word entropy, b) type token ratio and c) Zipf exponent. Timeseries are smoothed with a moving average window of $\pm$ 5 years, and averaged over media categories. Shaded region shows 95\% confidence interval of this average.}
		\label{fig:lexical-diversity-timeseries}
	\end{figure}

We analysed the Corpus of Historical American English (COHA), a balanced corpus with text samples from the 1810s to the 2000s categorised into news, magazines, fiction and non-fiction \cite{Davies2012Nov}. As discussed in the Methods section, we analysed text samples truncated to $N=2000$ words. We found a clear trend of rising lexical diversity since approximately 1900 as measured by word entropy, type token ratio and Zipf exponent (Figure \ref{fig:lexical-diversity-timeseries}). 

\begin{figure}[ht]
		\centering
		\includegraphics[width=1\textwidth]{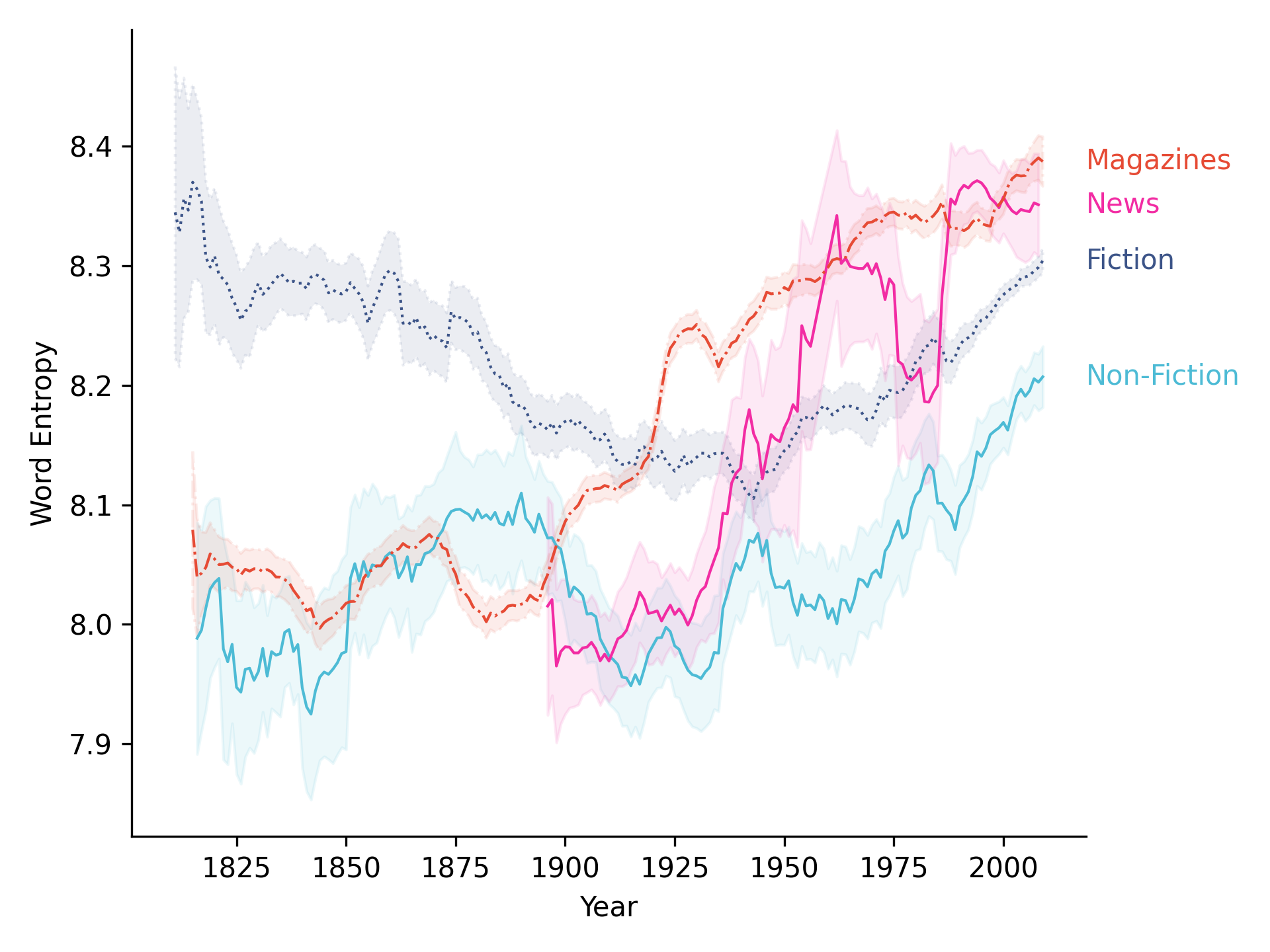}
		\caption{Timeseries of word entropy across media categories in the Corpus of Historical American English. For each media category, the timeseries was smoothed using an average over a window of $\pm$ 5 years. The shaded regions are 95\% confidence intervals of this average. All media categories show an upward trend in word entropy from 1900.}
		\label{fig:timeseries}
	\end{figure}

The trends in separate media categories follow the same pattern of rising lexical diversity as measured by word entropy (Figure \ref{fig:timeseries}). We analysed the timeseries of annual averages since 1900 for each media category (fiction, non-fiction, news, magazines) and lexical measure (word entropy, Zipf exponent, type token ratio) using Kwiatkowski–Phillips–Schmidt–Shin (KPSS) and Mann-Kendall (MK) tests on the annual median values (using the annual mean gives similar results). This gives a total of $4 \times 3 \times 2 = 24$ trend tests. All 24 tests show significant evidence of a trend at $p<0.05$ And 22 out of 24 tests show significant evidence of trends at $p<0.01$ (the tests for a trend in type token ratio in non-fiction had KPSS $p=0.015$ and MK $p=0.013$). Overall there is very strong evidence for a trend of rising lexical diversity in all media categories between 1900 and 2010. For full results and a deeper analysis, see the Supplementary Information. 

\subsection*{Higher Entropy in Short-form Media}

\begin{figure}[ht]
		\centering
		\includegraphics[width=1\textwidth]{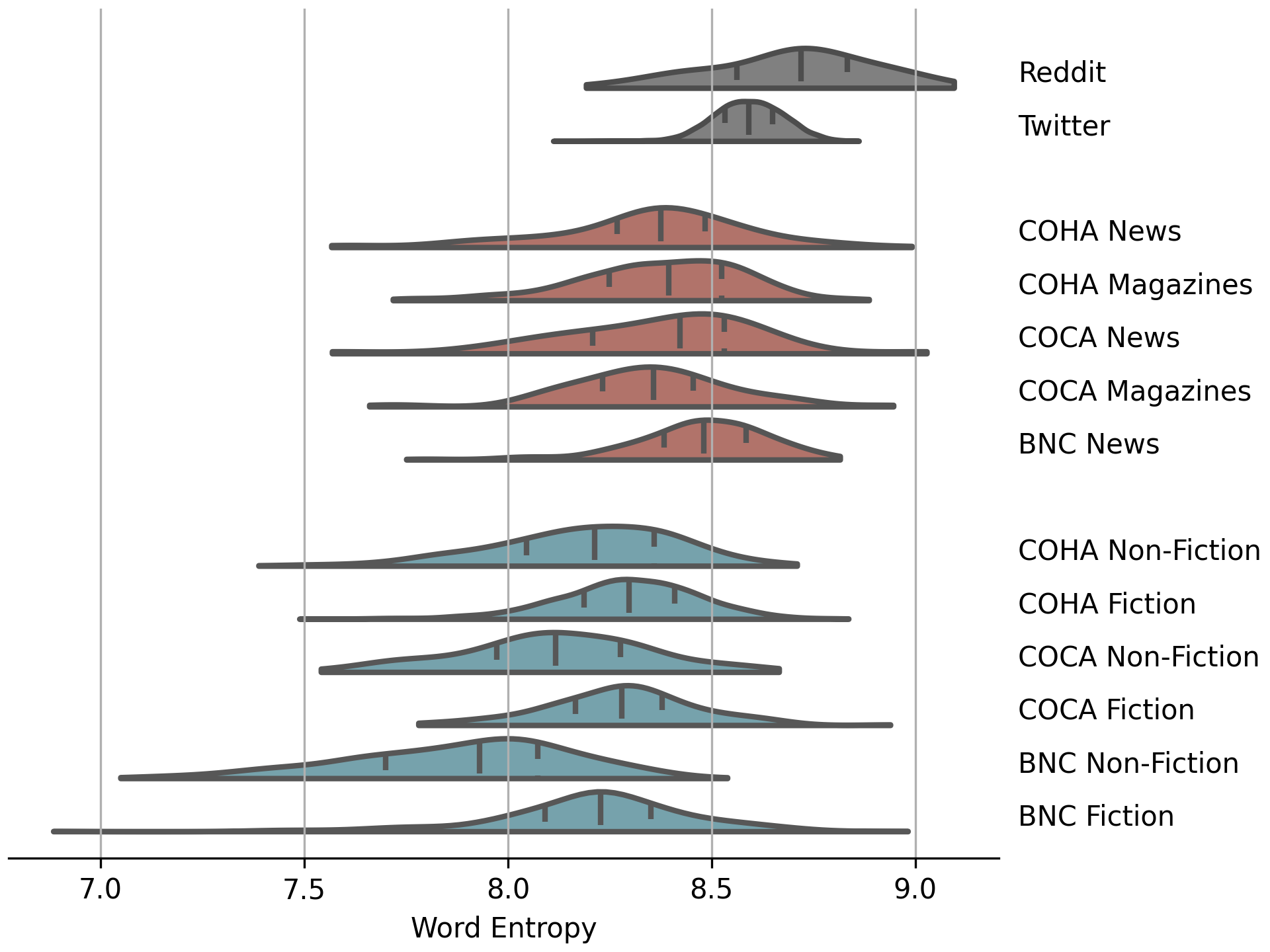}
		\caption{Word entropy of very short-form (social) media, short-form (news and magazines) and long-form (fiction and non-fiction) media. For each media category, distributions are kernel density estimates cut to the data range, with quartile positions shown. The COHA data was restricted to 2000-2007 to minimise the effect of historical changes.}
		\label{fig:media-categories}
	\end{figure}

The historical trend (Figure \ref{fig:timeseries}) suggests modern differences in entropy between media categories. However, we also know that short-form media has become especially prominent with the recent rise of online platforms for media distribution, such as social media, RSS feeds, and news platforms that present short headlines and snippets that link to long-form articles. To investigate these different media categories, we examined the Corpus of Contemporary American English (COCA) and the British National Corpus (BNC), as well as social media data from Twitter and Reddit. Figure \ref{fig:media-categories} shows the distribution of word entropy across different media categories. Within COHA (limited to 2000-2007), BNC, and COCA there were significant differences in all lexical measures across media categories (ANOVA tests $p<0.01$). Full statistical results are in the Supplementary Information. Overall, short-form media categories of news and magazines have higher entropy than long-form media, and social media feeds have the highest entropy of all. 

It should be noted that when analysing social media data we collated posts to create text samples with $N=2000$ words, to match the length of the other media type analyses. Combining posts will naturally lead to high entropy text, with fast switching of contexts and high novelty. This mirrors how people actually consume social media. Essentially, social media platforms generate high entropy information environments in the form of feeds of short messages from different users. This is not necessarily a linguistic change in how people generate English; it is a change in how people consume English text. 
 
\subsection*{Information Foraging in the Attention Economy}

The results are suggestive of a link between competition for attention and word entropy. To explain these results we generate a model of the attention economy based on information foraging. Foraging models relate the consumption of information items with some utility gain to the forager. To bridge utility rates to lexical measures, we borrow the idea of information signal entropy from Shannon \cite{shannon1948mathematical}: the entropy of a source of information is a function of the probability of seeing each symbol given the preceding symbols. For our purposes entropy can be thought of as a rate of information. If information foragers gain utility from information then, by definition, an increase in entropy, $h$, is associated with an increase in utility rate, $r$. This aligns with Zipf's principle of least effort \cite{zipf1949human, Ferreri.Cancho2005Mar}.

\begin{equation}
    h \propto r \,.
\end{equation}

Animal foragers modulate the selectivity of their diet in response to the environment, becoming more selective in times of abundance \cite{stephenskrebs1986foraging}. Why waste energy hunting difficult prey when there are plenty of easy calories around? Humans act in the same way when selecting information to consume \cite{pirolli1999information, simon1969designing}. We have all experienced situations where we do not have access to the internet, for example on a plane or train journey, and we become less selective in what we read or watch. 

This characterisation of attention corresponds to the prey choice model, which describes which types of prey are worth pursuing and consuming \cite{stephenskrebs1986foraging}.  And this has been applied to information foraging before \cite{pirolli1999information}. The derivation of the prey model followed here is exactly analogous to that found in the prey choice model in food foraging. Our contribution will come at the end of this section, where we extend the model to include media competition for attention. 

Assume an information forager searches a media environment and encounters information of types, $i$, at Poisson rates $\lambda_i$. If consumed, information provides a benefit $u_i$ in a handling time $t_i$, during which time the forager is not searching. Alternatively, the forager can choose to ignore information of a certain type and keep searching. The forager's choices to consume or ignore information determine the expected total time spent searching, $T_s$, and handling, $T_h$, information, as well as the total utility gain, $U$. Given these constraints, the forager aims to optimise the expected overall rate of utility of foraging given by

\begin{equation}
R_{media} = \dfrac{U}{T_s + T_h} \,.
\end{equation}

Here \emph{media} describes the forager's local environment, such as a media platform. Media platforms are analogous to foraging patches in optimal foraging theory. The forager's choices of which information types to consume can be described as an information diet, $D$. The total expected utility is $U = \sum_D \lambda_i u_i T_s$. Similarly the total expected handling time is $T_h = \sum_D \lambda_i t_i T_s$. Substituting in and cancelling $T_s$, we can write the expected utility rate given a diet

\begin{equation}
R_{media} = \dfrac{\sum_D \lambda_i u_i}{1 + \sum_D \lambda_i t_i} \,. \label{eqn:diet_rate}
\end{equation}

Consuming an information item carries an expected opportunity cost of not spending that item's handling time looking for other items, equal to $t_i R_{media}$, and an expected utility gain of $u_i$. To maximise expected utility rate a forager should therefore consume the item if the item utility rate, $r_i = \frac{u_i}{t_i}$, is greater than the overall media platform utility rate, $R_{media}$, 

\begin{equation}
    r_i \geq R_{media} \label{eqn:info_choice} \,.
\end{equation}

This diet threshold condition is a familiar result from foraging theory \cite{stephenskrebs1986foraging, macarthur1966optimal,pirolli1999information}. To find the optimal diet, item types can be ranked in order of $r_i$ and added to the diet one by one until this inequality fails \cite{macarthur1966optimal}. See the Supplementary Information for a more thorough derivation.

We can now ask which information types a forager should include in their diet, $D$, to maximise their
expected overall utility rate as a consequence of rising information prevalence, here $\lambda_i$. For items with $r_i < R_{media}$, increasing prevalence has no effect as these items are still not included in the diet. For items with $r_i \geq R_{media}$, increasing prevalence will mean more time spent handling these items and less time spent searching, so the overall media platform utility rate will increase,

\begin{equation}
    \frac{\partial R_{media}}{\partial \lambda_i} \geq 0 \quad \forall i  \label{eqn:r_rises} \,. 
\end{equation}

Combining this with the information diet criterion (Inequality \ref{eqn:info_choice}), increasing information prevalence increases the information utility rate required for diet inclusion: foragers become more selective when prey (or information) is abundant, analogous to the prey model in optimal foraging theory \cite{stephenskrebs1986foraging}. 

\begin{figure}[ht]
		\centering
		\includegraphics[width=1\textwidth]{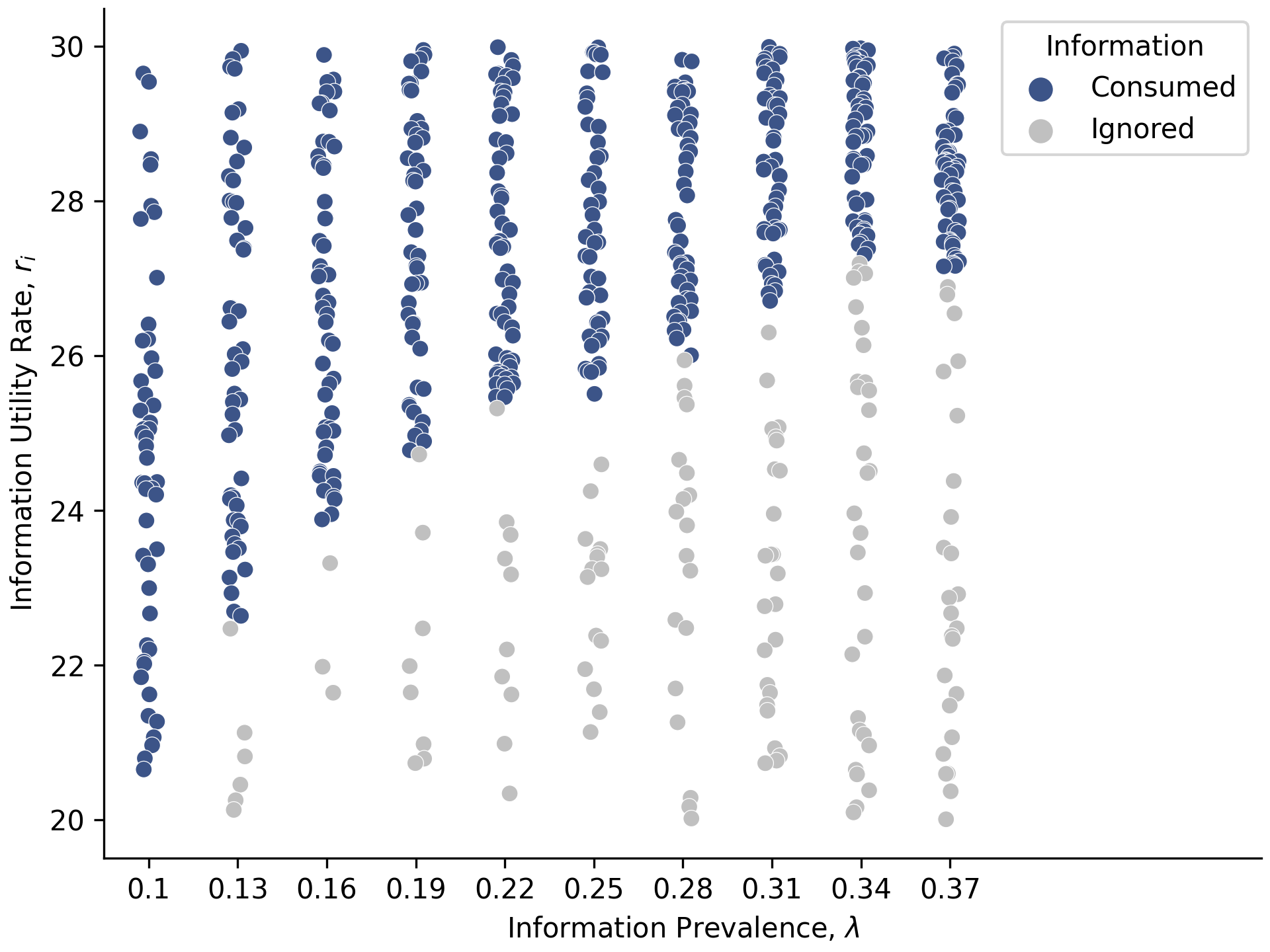}
		\caption{Simulation of information foraging in the attention economy. Information items are generated with random utility rates in quantities proportional to the information prevalence. Given the information environment, foragers only consume information items above a minimum information density (blue markers) in order to maximise their foraging rate.  Information that is not consumed has less chance of survival (grey markers). Overall the surviving information types have higher utility rates at higher information prevalence.}
		\label{fig:simulation}
	\end{figure}

We now extend traditional foraging theory to information co-evolution by asking how media producers respond to increasing selectivity among information foragers. By assuming there is some cost to media of producing more informative messages --- a standard assumption underlying Zipf's principle of least effort \cite{i2003least,zipf1949human} --- we conclude that an abundance of information creates an adaptive pressure that drives media producers to create information with a higher utility rate. A proxy for utility rate is information density, or word entropy. Figure \ref{fig:simulation} shows a simple simulation of this dynamic. 

\subsection*{Competition Between Media Platforms Drives Differences Between Short- and Long-form Media}

Information is distributed in media platforms (e.g., newspapers, magazines, books, Twitter, Reddit). The forager has to choose not only which information to consume within a media platform, but also which media platforms to visit. Analogous to the information choice model (Equation \ref{eqn:info_choice}): an optimal information forager will visit a media platform if the expected media utility rate is greater than the background utility rate from foraging in the overall environment (see Supplementary Information for the full model),

\begin{equation}
	R_{media} \geq R_{env} \label{eqn:patch_choice} \,.
\end{equation}

The utility rate of a media platform, $R_{media}$, is a summation over Poisson processes  (Equation \ref{eqn:diet_rate}). To simplify this, let $\bar{u}_m$ be the average utility of information items consumed in the media platform, $\bar{t}_m$ the average time spent consuming information items, and $\lambda_m$ the rate of encounter of any item in the diet. Equation \ref{eqn:diet_rate} then  becomes a variation of Holling's disc equation \cite{holling1959some} (full derivation in Supplementary Information)

\begin{equation}
R_{media} = \dfrac{{\lambda}_m\bar{u}_m}{1 + {\lambda}_m\bar{t}_m} \label{eqn:holling} \,.
\end{equation}

\begin{figure}[ht]
		\centering
		\includegraphics[width=1\textwidth]{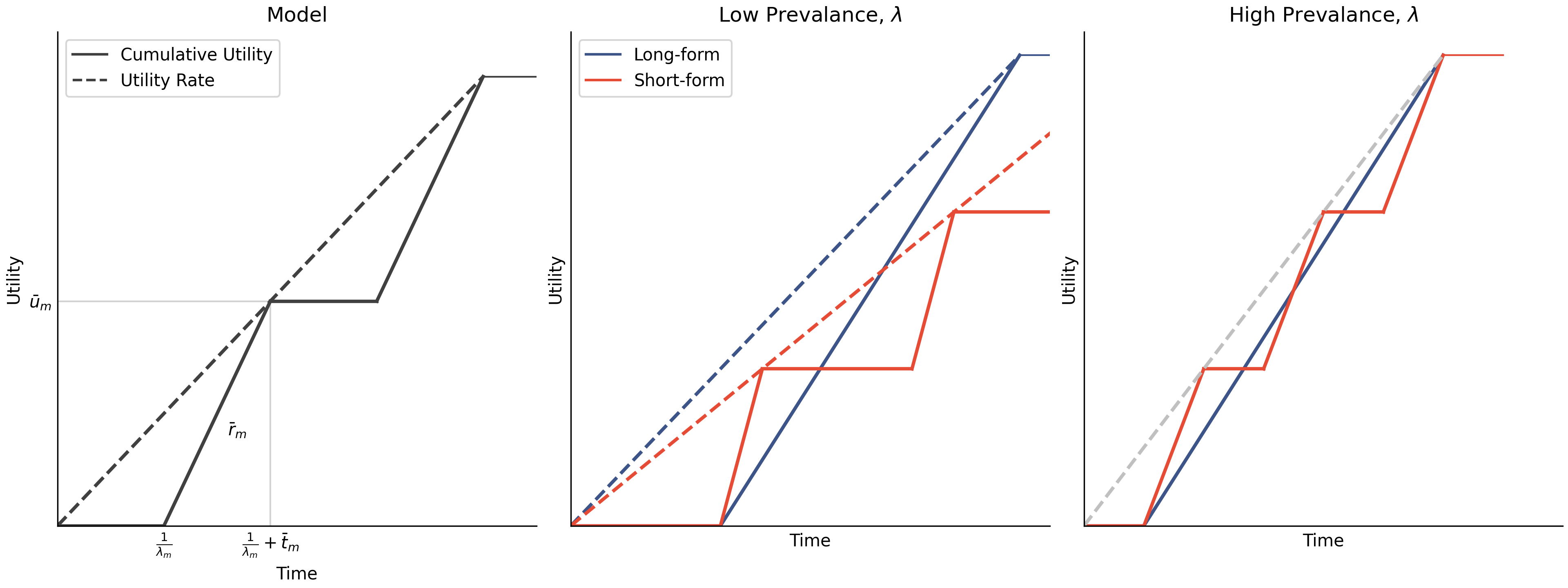}
		\caption{The media patch model. a) The expected utility rate of a media patch (dashed line) is determined by the time spent searching for (horizontal solid line) and consuming (diagonal solid line) information items. b) In a low prevalence environment long-form media has an advantage, although at low prevalence foragers are not very selective. c) At high prevalence less time is spent searching between item acquisition. To reach the same overall patch utility rate (dotted grey line), short-form media needs a higher information utility rate (gradient of the solid diagonal red line) than long-form media (gradient of the solid diagonal blue line). }
		\label{fig:patch_model}
	\end{figure}

This equation is visualised in Figure \ref{fig:patch_model} \textbf{a}.

The criteria for inclusion in an information forager's diet is then 

\begin{equation}
    \frac{1}{{\lambda}_m\bar{u}_m} + \frac{1}{\bar{r}_m} \leq \frac{1}{R_{env}} \label{eqn:media_patch_criteria}  \,.
\end{equation}

The inclusion of a media platform in the information diet is therefore determined by three properties of the information items that it contains and which would be included in the forager's information diet: the average utility (i.e. size) of a item, $\bar{u}_m$; the average item utility rate, $\bar{r}_m$; and the prevalence of items within the media platform, ${\lambda}_m$. 

Short-form media platforms such as news and magazines involve more time spent switching (and searching for) articles than long-form media platforms such as books. In order to reach the same overall media platform utility rate, $R_{media}$, short form media types need to have higher information utility rates (Figure \ref{fig:patch_model} \textbf{c}).  This creates a differential selective pressure on short- and long-form media producers. Given some $R_{env}$, the short-form media platform needs higher average information utility rates, $\bar{r}_m$, to be accepted in the forager's diet than the long-form media. The long-form media experiences a relaxed selective pressure on information utility rates because there is less time spent switching in these media platforms. This can describe the differences in the observed information utility rates in short- and long-form media as well as the trend towards increased information rates with increasing media prevalence.

\subsection*{Social Media}

\begin{figure}[ht]
		\centering
		\includegraphics[width=0.9\textwidth]{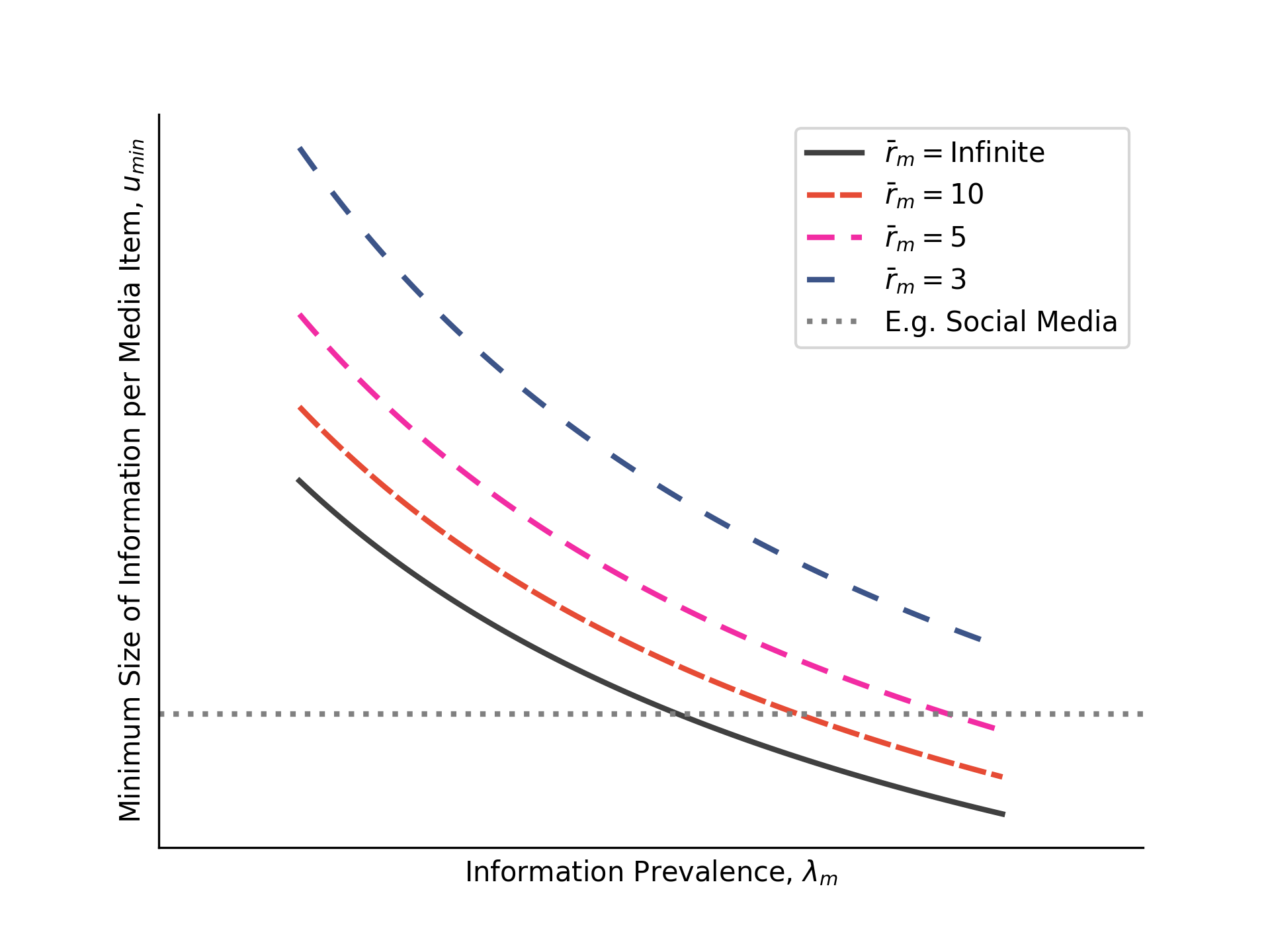}
		\caption{Minimum average information size, $u_{min}$, for media platform diet inclusion for varying levels of information prevalence, $\lambda_m$. Increasing average information utility rates, $\bar{r}_m$, can increase this limit only to a point. Very short-form media platforms like social media can only capture attention in a world with high information prevalence.}
		\label{fig:viability_social_media}
	\end{figure}

Inequality \ref{eqn:media_patch_criteria} includes a weaker condition for diet inclusion, $\frac{1}{\lambda_m \bar{u}_m} \leq \frac{1}{R_{env}}$. This indicates that information prevalence directly limits the minimal average size of information for diet inclusion. As information prevalence increases, foragers will tolerate media platforms with smaller and smaller information item sizes (Figure \ref{fig:viability_social_media}). More intuitively, Twitter only works in a world with instant messages --- few people would go to a library in order to check out a single Tweet. 

Finally, our model quantifies the selective forces acting to make media platforms more accessible. If a media platform reduces the expected search time between information encounters, $\frac{1}{\lambda_m}$, then they reduce the left hand side of Inequality \ref{eqn:media_patch_criteria} and become more competitive. This asymmetrically effects utility for short-form media, $\frac{1}{\lambda_m \bar{u}_m}$; for long-form media this term is already small. This could be an explanation for innovations towards minimising time spent searching in short-form media platforms such as infinite scroll and autoplay videos. 

\section*{Discussion}

We provide evidence that the word entropy of American English has increased over the 20th century. Furthermore, this change is marked by differences across different media categories, with the highest entropy levels found in the shortest media forms. Using a model of the attention economy based on information foraging, we show how a simple model of information selection can drive the observed changes.  The attention economy model explains two results: a rise in entropy as information becomes more abundant and a rise in preferences for information dense short-form media. 

Our findings offer an interesting contrast to the Linguistic Niche Hypothesis \cite{Lupyan2010Jan}, which predicts a loss of complex morphological forms in English due to the influence of second language learners. There is ample evidence that English is undergoing morphological simplification \cite{Michel2011Jan, Lieberman2007Oct, zhu2018modern}, and we might expect this to be associated with a decrease in word entropy (further explored in the Supplementary Information). Our findings show the opposite. Our claim is that the pressure towards information density overcomes the effect of reduced word entropy through linguistic simplification. However it may be that a reduction in morphological complexity and a rise in information entropy are related --- in attention markets people may be attracted to both simplicity \cite{hills2015recent} and novelty. Specifically, a loss in morphological complexity may be driven by a pressure towards simplicity and a reduction in the repetition of more difficult to process linguistic forms. That is, the features of the attention economy that drive rising entropy may also drive reduced morphological complexity. 

Language evolution has been shown to follow a number of principles governed by human psychology. These principles have, for example, included features of biological and cultural evolution \cite{smith2008cultural, christiansen2008language}, learning \cite{hills2015recent, christiansen2008language, Lupyan2010Jan}, cooling by expansion \cite{petersen2012languages}, word formation and distribution \cite{i2003least}, and the decay of morphological complexity \cite{Lupyan2010Jan, Lieberman2007Oct}. Our results extend the psychological consequences on language evolution to word entropy in response to information abundance. 

Considering people as information foragers, our model describes observed empirical changes in word entropy of English over time and both within and between media categories in response to increasing information abundance. Empirical findings support the idea that people's attention is attracted to high entropy and high complexity information \cite{itti2009bayesian, radach2003eye}. Our analysis of historical data shows the entropy of information markets respond predictably to increased competition. The attention economy model offers a simple explanation: humans are, within limits, information rate maximisers responding to rising information abundance and media producers adapt their content to compete for more limited attention. 

Humans choices are based on more than entropy. For example, humans respond to social cues and risk \cite{hills2019dark} just as animals consider factors other than calorie rate such as macro-nutrient content and predators when foraging for food \cite{stephenskrebs1986foraging}. Moreover, information producers are not only interested in capturing attention, but also in influence \cite{chen2014economic, evans2020economics}. Nonetheless, just as animal foraging models have been shown to predict human behaviour in a variety of domains \cite{winterhalder1986diet, pirolli1999information,pirolli2009elementary, fu2007snif,hills2012optimal}, our analyses suggests these models also extend to the shape of information evolution and cultural history, just as the co-evolutionary arguments of Darwin might have predicted \cite{darwin2011various}. 

\subsection*{Data Availability}

All data generated following analysis of text samples is available at \url{https://github.com/chasmani/PUBLIC-the-rising-entropy-of-english-in-the-attention-economy}.

The text corpora data is not included in the public repository for copyright and size reasons. They are available:

\begin{itemize}
    \item COHA and COCA. \url{https://www.corpusdata.org/}
    \item BNC. \url{http://www.natcorp.ox.ac.uk/}
    \item Twitter dataset. \url{https://www.kaggle.com/kazanova/sentiment140}
    \item Reddit dataset. This was collected from Reddit's API on 23rd March 2023. We cannot provide the data for copyright reasons. \url{https://www.reddit.com/.json}. 
\end{itemize}

\subsection*{Code Availability}

All code used to generate figures and analysis is available at \url{https://github.com/chasmani/PUBLIC-the-rising-entropy-of-english-in-the-attention-economy}.

\bibliographystyle{unsrt}
\bibliography{refs}

\begin{thebibliography}{10}

\bibitem{zipf1949human}
George~Kingsley Zipf.
\newblock {\em Human behavior and the principle of least effort: An
  introduction to human ecology}.
\newblock Hafner, 1949.

\bibitem{Bentz2015Jun}
Christian Bentz, Annemarie Verkerk, Douwe Kiela, Felix Hill, and Paula Buttery.
\newblock {Adaptive Communication: Languages with More Non-Native Speakers Tend
  to Have Fewer Word Forms}.
\newblock {\em PLoS One}, 10(6):e0128254, Jun 2015.

\bibitem{baixeries2013evolution}
Jaume Baixeries, Brita Elvev{\aa}g, and Ramon Ferrer-i Cancho.
\newblock The evolution of the exponent of zipf's law in language ontogeny.
\newblock {\em PloS one}, 8(3):e53227, 2013.

\bibitem{Ferreri.Cancho2005Mar}
R.~Ferrer~i. Cancho.
\newblock {The variation of Zipf{'}s law in human language}.
\newblock {\em Eur. Phys. J. B}, 44(2):249--257, March 2005.

\bibitem{hills2019dark}
Thomas~T Hills.
\newblock The dark side of information proliferation.
\newblock {\em Perspectives on Psychological Science}, 14(3):323--330, 2019.

\bibitem{simon1955behavioral}
Herbert~A Simon.
\newblock A behavioral model of rational choice.
\newblock {\em The quarterly journal of economics}, 69(1):99--118, 1955.

\bibitem{evans2020economics}
David~S Evans.
\newblock The economics of attention markets.
\newblock {\em Available at SSRN 3044858}, 2020.

\bibitem{ciampaglia2015production}
Giovanni~Luca Ciampaglia, Alessandro Flammini, and Filippo Menczer.
\newblock The production of information in the attention economy.
\newblock {\em Scientific reports}, 5(1):1--6, 2015.

\bibitem{terranova2012attention}
Tiziana Terranova.
\newblock Attention, economy and the brain.
\newblock {\em Culture Machine}, 13, 2012.

\bibitem{Lorenz-Spreen2019Apr}
Philipp Lorenz-Spreen, Bjarke~M{\o}rch M{\o}nsted, Philipp
  H{\ifmmode\ddot{o}\else\"{o}\fi}vel, and Sune Lehmann.
\newblock {Accelerating dynamics of collective attention}.
\newblock {\em Nat. Commun.}, 10(1759):1--9, Apr 2019.

\bibitem{pirolli1999information}
Peter Pirolli and Stuart Card.
\newblock Information foraging.
\newblock {\em Psychological review}, 106(4):643, 1999.

\bibitem{sandstrom1994optimal}
Pamela~Effrein Sandstrom.
\newblock An optimal foraging approach to information seeking and use.
\newblock {\em The library quarterly}, 64(4):414--449, 1994.

\bibitem{pirolli2009information}
Peter~LT Pirolli.
\newblock {\em Information foraging theory: Adaptive interaction with
  information}.
\newblock Oxford University Press, 2009.

\bibitem{lawrance2010programmers}
Joseph Lawrance, Christopher Bogart, Margaret Burnett, Rachel Bellamy, Kyle
  Rector, and Scott~D Fleming.
\newblock How programmers debug, revisited: An information foraging theory
  perspective.
\newblock {\em IEEE Transactions on Software Engineering}, 39(2):197--215,
  2010.

\bibitem{lawrance2010reactive}
Joseph Lawrance, Margaret Burnett, Rachel Bellamy, Christopher Bogart, and
  Calvin Swart.
\newblock Reactive information foraging for evolving goals.
\newblock In {\em Proceedings of the SIGCHI Conference on Human Factors in
  Computing Systems}, pages 25--34, 2010.

\bibitem{piorkowski2013whats}
David~J Piorkowski, Scott~D Fleming, Irwin Kwan, Margaret~M Burnett,
  Christopher Scaffidi, Rachel~KE Bellamy, and Joshua Jordahl.
\newblock The whats and hows of programmers' foraging diets.
\newblock In {\em Proceedings of the SIGCHI Conference on Human Factors in
  Computing Systems}, pages 3063--3072, 2013.

\bibitem{bhowmik2015optimal}
Tanmay Bhowmik, Nan Niu, Wentao Wang, Jing-Ru~C Cheng, Ling Li, and Xiongfei
  Cao.
\newblock Optimal group size for software change tasks: A social information
  foraging perspective.
\newblock {\em IEEE transactions on cybernetics}, 46(8):1784--1795, 2015.

\bibitem{hills2015exploration}
Thomas~T Hills, Peter~M Todd, David Lazer, A~David Redish, Iain~D Couzin,
  Cognitive Search~Research Group, et~al.
\newblock Exploration versus exploitation in space, mind, and society.
\newblock {\em Trends in Cognitive Sciences}, 19(1):46--54, 2015.

\bibitem{hills2006animal}
Thomas~T Hills.
\newblock Animal foraging and the evolution of goal-directed cognition.
\newblock {\em Cognitive science}, 30(1):3--41, 2006.

\bibitem{todd2020foraging}
Peter~M Todd and Thomas~T Hills.
\newblock Foraging in mind.
\newblock {\em Current Directions in Psychological Science}, 29(3):309--315,
  2020.

\bibitem{Davies2012Nov}
Mark Davies.
\newblock {Expanding horizons in historical linguistics with the 400-million
  word Corpus of Historical American English}.
\newblock {\em Edinburgh University Press 22 George Square, Edinburgh EH8 9LF
  UK}, Nov 2012.

\bibitem{davies2009385+}
Mark Davies.
\newblock The 385+ million word corpus of contemporary american english
  (1990--2008+): Design, architecture, and linguistic insights.
\newblock {\em International journal of corpus linguistics}, 14(2):159--190,
  2009.

\bibitem{ByLouBurnard2007Jan}
Edited By~Lou~Burnard.
\newblock {Reference Guide for the British National Corpus (XML Edition)}, Jan
  2007.
\newblock [Online; accessed 18. Mar. 2021].

\bibitem{gerlach2020standardized}
Martin Gerlach and Francesc Font-Clos.
\newblock A standardized project gutenberg corpus for statistical analysis of
  natural language and quantitative linguistics.
\newblock {\em Entropy}, 22(1):126, 2020.

\bibitem{rudnicka2018variation}
Karolina Rudnicka.
\newblock Variation of sentence length across time and genre.
\newblock {\em Diachronic corpora, genre, and language change}, pages 220--240,
  2018.

\bibitem{bird2009natural}
Steven Bird, Ewan Klein, and Edward Loper.
\newblock {\em Natural language processing with Python: analyzing text with the
  natural language toolkit}.
\newblock " O'Reilly Media, Inc.", 2009.

\bibitem{go2009twitter}
Alec Go, Richa Bhayani, and Lei Huang.
\newblock Twitter sentiment classification using distant supervision.
\newblock {\em CS224N project report, Stanford}, 1(12):2009, 2009.

\bibitem{bentz2017entropy}
Christian Bentz, Dimitrios Alikaniotis, Michael Cysouw, and Ramon Ferrer-i
  Cancho.
\newblock The entropy of words—learnability and expressivity across more than
  1000 languages.
\newblock {\em Entropy}, 19(6):275, 2017.

\bibitem{clauset2009power}
Aaron Clauset, Cosma~Rohilla Shalizi, and Mark~EJ Newman.
\newblock Power-law distributions in empirical data.
\newblock {\em SIAM review}, 51(4):661--703, 2009.

\bibitem{pilgrim2020bias}
Charlie Pilgrim and Thomas~T Hills.
\newblock Bias in zipf's law estimators.
\newblock {\em arXiv preprint arXiv:2008.00903}, 2020.

\bibitem{pilgrim2021piecewise}
Charlie Pilgrim.
\newblock Piecewise-regression (aka segmented regression) in python.
\newblock {\em Journal of Open Source Software}, 6(68):3859, 2021.

\bibitem{seabold2010statsmodels}
Skipper Seabold and Josef Perktold.
\newblock Statsmodels: Econometric and statistical modeling with python.
\newblock In {\em Proceedings of the 9th Python in Science Conference},
  volume~57, page~61. Austin, TX, 2010.

\bibitem{hussain2019pymannkendall}
Md~Manjurul Hussain and Ishtiak Mahmud.
\newblock pymannkendall: a python package for non parametric mann kendall
  family of trend tests.
\newblock {\em Journal of Open Source Software}, 4(39):1556, 2019.

\bibitem{sumner2010magazine}
David~E Sumner.
\newblock {\em The magazine century: American magazines since 1900}, volume~9.
\newblock Peter Lang, 2010.

\bibitem{shannon1948mathematical}
Claude~E Shannon.
\newblock A mathematical theory of communication.
\newblock {\em The Bell system technical journal}, 27(3):379--423, 1948.

\bibitem{stephenskrebs1986foraging}
David~W Stephens and John~R Krebs.
\newblock {\em Foraging theory}, volume~1.
\newblock Princeton University Press, 1986.

\bibitem{simon1969designing}
Herbert~A Simon.
\newblock Designing organizations for an information-rich world.
\newblock {\em Brookings Institute Lecture}, 1969.

\bibitem{macarthur1966optimal}
Robert~H MacArthur and Eric~R Pianka.
\newblock On optimal use of a patchy environment.
\newblock {\em The American Naturalist}, 100(916):603--609, 1966.

\bibitem{i2003least}
Ramon~Ferrer i~Cancho and Ricard~V Sol{\'e}.
\newblock Least effort and the origins of scaling in human language.
\newblock {\em Proceedings of the National Academy of Sciences},
  100(3):788--791, 2003.

\bibitem{holling1959some}
Crawford~S Holling.
\newblock Some characteristics of simple types of predation and parasitism.
\newblock {\em Canadian entomologist}, 91(7):385--398, 1959.

\bibitem{Lupyan2010Jan}
Gary Lupyan and Rick Dale.
\newblock {Language Structure Is Partly Determined by Social Structure}.
\newblock {\em PLoS One}, 5(1):e8559, Jan 2010.

\bibitem{Michel2011Jan}
Jean-Baptiste Michel, Yuan~Kui Shen, Aviva~Presser Aiden, Adrian Veres,
  Matthew~K. Gray, {}, Joseph~P. Pickett, Dale Hoiberg, Dan Clancy, Peter
  Norvig, Jon Orwant, Steven Pinker, Martin~A. Nowak, and Erez~Lieberman Aiden.
\newblock {Quantitative Analysis of Culture Using Millions of Digitized Books}.
\newblock {\em Science}, 331(6014):176--182, Jan 2011.

\bibitem{Lieberman2007Oct}
Erez Lieberman, Jean-Baptiste Michel, Joe Jackson, Tina Tang, and Martin~A.
  Nowak.
\newblock {Quantifying the evolutionary dynamics of language}.
\newblock {\em Nature}, 449(7163):713--716, Oct 2007.

\bibitem{zhu2018modern}
Haoran Zhu and Lei Lei.
\newblock Is modern english becoming less inflectionally diversified? evidence
  from entropy-based algorithm.
\newblock {\em Lingua}, 216:10--27, 2018.

\bibitem{hills2015recent}
Thomas~T Hills and James~S Adelman.
\newblock Recent evolution of learnability in american english from 1800 to
  2000.
\newblock {\em Cognition}, 143:87--92, 2015.

\bibitem{smith2008cultural}
Kenny Smith and Simon Kirby.
\newblock Cultural evolution: implications for understanding the human language
  faculty and its evolution.
\newblock {\em Philosophical Transactions of the Royal Society B: Biological
  Sciences}, 363(1509):3591--3603, 2008.

\bibitem{christiansen2008language}
Morten~H Christiansen and Nick Chater.
\newblock Language as shaped by the brain.
\newblock {\em Behav Brain Sci}, 31(5):489--509, 2008.

\bibitem{petersen2012languages}
Alexander~M Petersen, Joel~N Tenenbaum, Shlomo Havlin, H~Eugene Stanley, and
  Matja{\v{z}} Perc.
\newblock Languages cool as they expand: Allometric scaling and the decreasing
  need for new words.
\newblock {\em Scientific reports}, 2(1):1--10, 2012.

\bibitem{itti2009bayesian}
Laurent Itti and Pierre Baldi.
\newblock Bayesian surprise attracts human attention.
\newblock {\em Vision research}, 49(10):1295--1306, 2009.

\bibitem{radach2003eye}
Ralph Radach, Stefanie Lemmer, Christian Vorstius, Dieter Heller, and Karina
  Radach.
\newblock Eye movements in the processing of print advertisements.
\newblock In {\em The Mind's Eye}, pages 609--632. Elsevier, 2003.

\bibitem{chen2014economic}
Jianqing Chen and Jan Stallaert.
\newblock An economic analysis of online advertising using behavioral
  targeting.
\newblock {\em Mis Quarterly}, 38(2):429--A7, 2014.

\bibitem{winterhalder1986diet}
Bruce Winterhalder.
\newblock Diet choice, risk, and food sharing in a stochastic environment.
\newblock {\em Journal of anthropological archaeology}, 5(4):369--392, 1986.

\bibitem{pirolli2009elementary}
Peter Pirolli.
\newblock An elementary social information foraging model.
\newblock In {\em Proceedings of the SIGCHI conference on human factors in
  computing systems}, pages 605--614, 2009.

\bibitem{fu2007snif}
Wai-Tat Fu and Peter Pirolli.
\newblock Snif-act: A cognitive model of user navigation on the world wide web.
\newblock {\em Human--Computer Interaction}, 22(4):355--412, 2007.

\bibitem{hills2012optimal}
Thomas~T Hills, Michael~N Jones, and Peter~M Todd.
\newblock Optimal foraging in semantic memory.
\newblock {\em Psychological review}, 119(2):431, 2012.

\bibitem{darwin2011various}
Charles Darwin.
\newblock {\em On the Various Contrivances by which Orchids are Fertilized by
  Insects (1862)}.
\newblock University of Chicago Press, 2011.

\bibitem{kettunen2014can}
Kimmo Kettunen.
\newblock Can type-token ratio be used to show morphological complexity of
  languages?
\newblock {\em Journal of Quantitative Linguistics}, 21(3):223--245, 2014.

\bibitem{deutscher2009overall}
Guy Deutscher et~al.
\newblock " overall complexity": a wild goose chase?
\newblock 2009.

\bibitem{sampson2009linguistic}
Geoffrey Sampson.
\newblock A linguistic axiom challenged.
\newblock {\em Language complexity as an evolving variable}, 2:18, 2009.

\bibitem{Charnov1976Apr}
Eric~L. Charnov.
\newblock {Optimal foraging, the marginal value theorem}.
\newblock {\em Theor. Popul. Biol.}, 9(2):129--136, Apr 1976.

\bibitem{bettinger2016marginal}
Robert~L Bettinger and Mark~N Grote.
\newblock Marginal value theorem, patch choice, and human foraging response in
  varying environments.
\newblock {\em Journal of Anthropological Archaeology}, 42:79--87, 2016.

\bibitem{po2003news}
Horst P\"{o}ttker.
\newblock News and its communicative quality: the inverted pyramid—when and
  why did it appear?
\newblock {\em Journalism Studies}, 4(4):501--511, 2003.

\bibitem{gallager2012discrete}
Robert~G Gallager.
\newblock {\em Discrete stochastic processes}, volume 321.
\newblock Springer Science \& Business Media, 2012.

\end{thebibliography}

\subsection*{Acknowledgements}

The study was funded by the EPSRC grant for the Mathematics for Real-World Systems CDT at Warwick (grant number EP/L015374/1).  T.T.H. was supported on this work by the Royal Society Wolfson Research Merit Award (WM160074) and a Fellowship from the Alan Turing Institute, which is funded by EPSRC (grant number EP/N510129/1). 

\subsection*{Author Contributions}

C.P. and T.T.H. conceived and developed the presented idea. C.P. developed the mathematical model and carried out data analysis, with guidance from T.T.H.  C.P. took the lead in writing the manuscript, T.T.H. and W.G. gave revisions and feedback. T.T.H. supervised the project throughout. All authors reviewed the results and approved the final version of the manuscript. 

\subsection*{Competing Interests}

The authors declare no competing interests. 

\subsection*{Materials and Correspondance}

Direct correspondence to Charlie Pilgrim at c.pilgrim@ucl.ac.uk.

\pagebreak

\section{Supplementary Information --- Linguistic Niche Hypothesis}

The finding in the main paper of word entropy, and lexical diversity, rising in American English is the opposite of what might be predicted by the Linguistic Niche Hypothesis. That hypothesis makes predictions about the complexity of language morphology (e.g. I ate, la casita) and syntax (e.g. I did eat, la pequeña casa), with the assumption that complexity is balanced between the two. The Linguistic Niche Hypothesis \cite{Lupyan2010Jan} suggests that languages in large, spread out social systems tend to have simpler morphological forms, with the grammatical work instead being done through syntax  \cite{Lupyan2010Jan}. The hypothesised mechanism for this is that second language learners prefer simpler forms so that complex morphological forms disappear over time \cite{Lupyan2010Jan}. A global lingua franca like English should therefore be undergoing morphological simplification, and evidence does suggest that this is the case with the regularisation of English past tense verbs \cite{Michel2011Jan, Lieberman2007Oct} and a loss of inflectional diversity \cite{zhu2018modern}. Further work suggests that this morphological simplification should correlate with a reduction in lexical diversity as measured by type token ratio \cite{Bentz2015Jun, kettunen2014can} (or word entropy) --- complex morphological forms are non-repetitive (many unique word types per word token) whilst syntactic grammatical modifiers are repetitive (few unique word types per word token). We find that lexical diversity is instead rising in American English. We suggest some possible explanations:

\begin{enumerate}
  \item English morphology is overall becoming more complex, against the Linguistic Niche Hypothesis.
  \item English morphology is becoming simpler without an increase in syntactic complexity. This would be a further refutation of the already beleaguered \cite{deutscher2009overall, sampson2009linguistic} equicomplexity assumption, which states that mature languages have broadly equal grammatical complexity, balanced between morphology and syntax.       
  \item Lexical diversity (and Type Token Ratio) is not a good measure of morphological complexity.  The increase in lexical diversity is instead driven by more concise information and a wider, and faster switching of, contexts in written media. 
\end{enumerate}

The third option here aligns well with the ideas in the main paper, and is in our opinion at least partly responsible. If people are drawn towards higher utility rate information then that could drive English to be more concise and to switch contexts more quickly.

\section{Supplementary Information - Historical Analysis of US Magazine Publishing}

\begin{figure}[ht]
		\centering
		\includegraphics[width=1\textwidth]{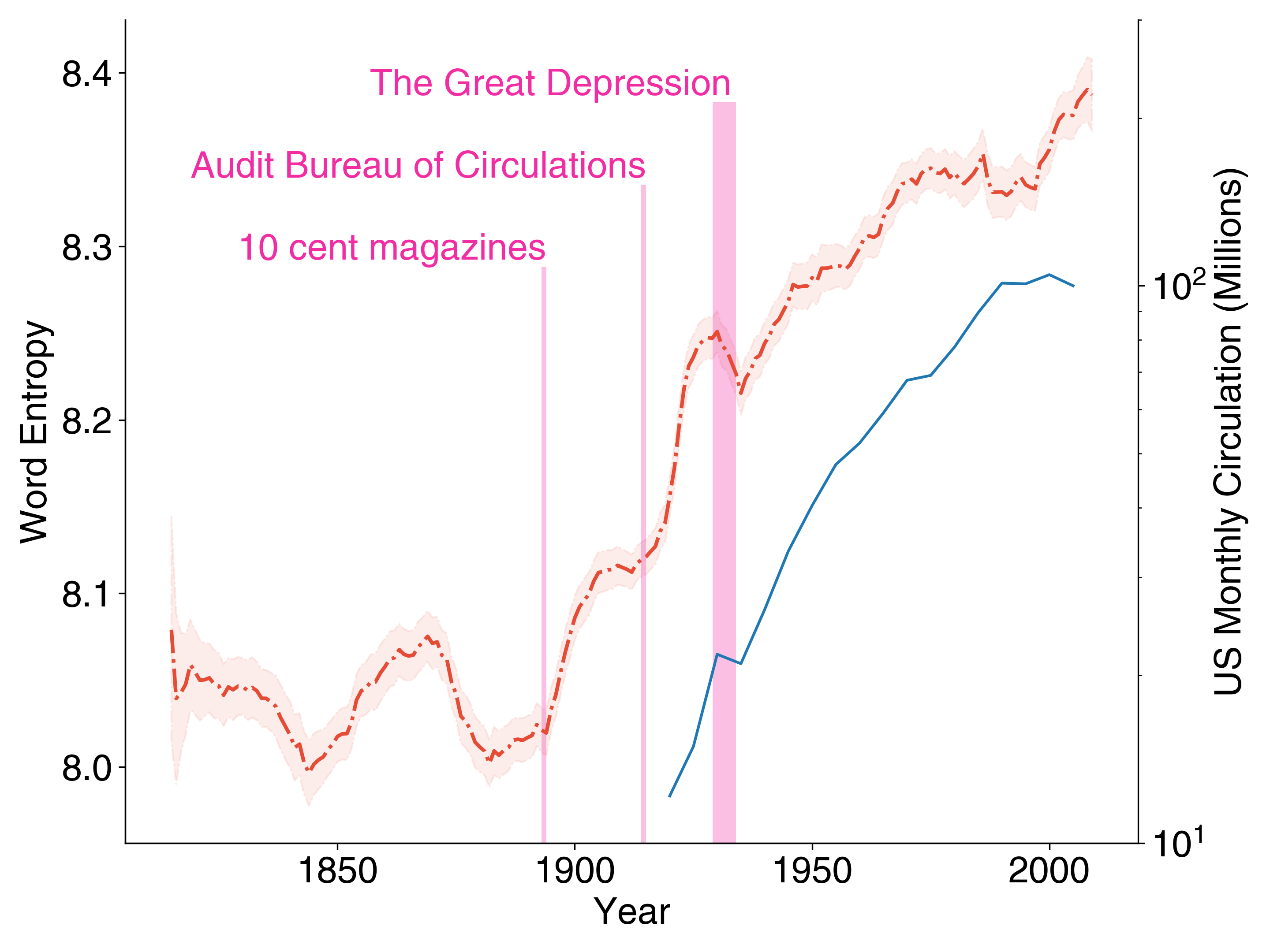}
		\caption{Historical analysis of word entropy in magazines (red dotted, timeseries calculated as in previous figure) with key events (pink) and US Monthly Magazine circulation as reported by the Audit Bureau of Circulations (purple).}
		\label{fig:historical-magazines}
	\end{figure}

As a case study we investigated the history of magazine publishing in America. Figure \ref{fig:historical-magazines} shows the historical trend in COHA magazine word entropy alongside magazine circulation figures and important events.  Magazine publishers are in a two-sided market where they sell magazines to consumers and attention to advertisers \cite{evans2020economics}, with the majority of revenue from selling attention \cite{sumner2010magazine}.  This wasn't always the case in the US --- prior to the 1890s most magazine revenue was from sales, with advertising considered undesirable \cite{sumner2010magazine}. Towards the late 19th century a combination of rapidly decreasing printing costs, growth in the literate population, discounts from the US postal service and the ability to target adverts to a niche readership led to a new business model to emerge \cite{sumner2010magazine}. This new model involved selling magazines lower than the price of production, which increased circulation so that those costs could be recouped by advertising revenue \cite{sumner2010magazine}. Before 1893, most magazines sold for 25 cents --- until a price war led to the magazines McClure's, Munsey's and Cosmopolitan dropping their prices to 10 cents and subsequently enjoying rises in circulation and advertising revenue \cite{sumner2010magazine}. The 10 cent magazines contributed to a tripling in total magazine readership from 1890 in 1905 \cite{sumner2010magazine}, and there was a huge jump in word entropy in the same period (Figure \ref{fig:historical-magazines}). 

The Audit Bureau of Circulation was created by advertisers in 1914 \cite{sumner2010magazine} to more accurately measure magazine readership numbers. This quantification of attention further increased pressure on magazine publishers to improve their circulation numbers in order to sell advertising. Other changes included moving advertisements from the back of the magazine to alongside the main content --- a move that forced copywriters to improve the appeal of the content through adding color and improving graphics \cite{sumner2010magazine}. 

Word entropy continues to rise throughout the 20th century alongside magazine circulation, with a Pearson's correlation coefficient r$=0.91$ ($p < 0.001$), although both rise over time so that confounding factors are not ruled out (Figure \ref{fig:historical-magazines}). After the 1890s, the biggest drop in word entropy was during the great depression when magazine circulation also fell. There is a suggestion in the data that things change around the year 2000, as magazine circulation drops but word entropy continues to rise. The rise of digital media around this time is perhaps the biggest change in publishing since the printing press so we would not expect the same trends to necessarily continue --- and digital media represents a new competitive pressure. 

\section{Supplementary Information --- Prey Choice Model Derivation}

In the main paper we justify the prey choice algorithm using an argument that considers the opportunity cost of spending time handling a prey versus searching in the environment. Here we derive the same result more rigorously. This is a completely analogous derivation as found in optimal foraging theory \cite{stephenskrebs1986foraging}. As in the main paper, we have information types, $i$, that are encountered with rates $\lambda_i$ while searching. Each information item, if consumed, provides a benefit $u_i$ in a handling time $t_i$, during which the forager is not searching for other items. 

In the main text, a media patch expected utility rate is given by,

\begin{equation}
R_{media} = \dfrac{\sum_D \lambda_i u_i}{1 + \sum_D \lambda_i t_i} \,. \label{eqn:diet_rate_SI}
\end{equation}

This assumes that information types are either in the diet, $D$, in which case they are always consumed upon encounter, or alternatively the items are not in the diet and never consumed. We can generalise this so that forager's have some probability of consuming an information type upon encounter, $p_i$, 

\begin{equation}
R_{media} = \dfrac{\sum \lambda_i u_i p_i }{1 + \sum \lambda_i t_i p_i}  \label{eqn:diet_rate_SI} \,.
\end{equation}

The forager can choose the probability of paying attention to each information type, and a forager's strategy can be defined as a vector 
$\textbf{p} = [p_1, p_2, . . . , p_n]$. These choices are independent. To find the strategy that gives the maximum utility rate we can consider each of these choices, $p_j$, independently. To find the best strategy we separate $p_j$ from the summations and differentiate

\begin{equation}
\frac{\partial R_{media}}{\partial p_j} = \dfrac{\lambda_j u_j (1 + p_j \lambda_j t_j + \sum_{i \neq j} p_i \lambda_i t_i) - \lambda_j t_j (p_j \lambda_j u_j + \sum_{i \neq j} p_i \lambda_i u_i)}{(1 + p_j \lambda_j t_j + \sum_{i \neq j} p_i \lambda_i t_i)^2} \,.
\end{equation}

Cancelling like terms

\begin{equation}
\frac{\partial R_{media}}{\partial p_j} = \dfrac{\lambda_j u_j (1 + \sum_{i \neq j} p_i \lambda_i t_i) - \lambda_j t_j (\sum_{i \neq j} p_i \lambda_i u_i)}{(1 + p_j \lambda_j t_j + \sum_{i \neq j} p_i \lambda_i t_i)^2} \,.
\end{equation}

The sign of this does not depend on $p_j$. So if $\frac{\partial R}{\partial p_j} > 0$, $R_{media}$ will be maximised with $p_j=1$, and otherwise with $p_j=0$. The condition for $p_j=1$ is 

\begin{equation}
\frac{u_j}{t_j} > \dfrac{\sum_{i \neq j} p_i \lambda_i u_i}{1 + \sum_{i \neq j} p_i \lambda_i t_i)} \label{eqn:diet_confition_full} \,.
\end{equation}

The right hand side is the total expected rate of utility for all items except for item $j$, $R_{\neg j}$. The item should be included in the diet if the utility rate of the item, $r_i = \frac{u_j}{t_j}$, is greater than the overall rate of foraging without the item. 

\begin{equation}
r_j \geq R_{\neg j} \label{eqn:diet_condition} \,.
\end{equation}

This is equivalent to the diet inclusion criteria given in the main paper. To find the optimal diet, one can add items in order of their utility rate until the inequality fails. 

\section{Supplementary Information --- Patch Choice Model and Non Constant Patches}

The patch choice model considered in the main paper is analogous to the information choice model. Patches of each type are randomly encountered in the environment and encountered as a Poisson processes with rates $\lambda_{media}$. We also assume that patches have a constant expected rate of utility, $R_{media}$, and some finite time, $T_{media}$ until the rate drops to zero, which gives each patch a total utility, $U_{media}$. Foragers can choose to either consume or ignore a patch upon encountering it. This model is identical to the information choice model so that we can follow that derivation and jump to the conclusion that a patch will be included in the diet if the patch utility rate is greater than or equal to the overall rate of foraging in the environment, $R_{media} \geq R_{env}$.  

Information patches in the real world have non-constant utility rates. Commonly patch marginal utility will decrease with time \cite{stephenskrebs1986foraging, Charnov1976Apr}. This can happen as finite prey are consumed \cite{bettinger2016marginal, stephenskrebs1986foraging}. For example, within a patch an optimal forager will consume the most profitable items first if they can, which then makes those items more scarce and reduces the overall utility rate in the patch as time goes on \cite{bettinger2016marginal}. Examples are collecting raspberries from a bush, or checking your email. Information items themselves may degrade while being consumed, for example news articles often follow an inverted pyramid structure where the most important information is presented first, with extra paragraphs adding marginally diminishing extra information \cite{po2003news}. Magazines, fiction and non-fiction have their own styles and utility curves. Overall we can say that utility rates in patches, and information, are not constant. 

An optimal forager now has to choose both which patches to consume and how long to spend in those patches. This problem was solved by Charnov's marginal value theorem \cite{Charnov1976Apr}, which we derive here in the context of information items. We follow the model and derivation given by Stephens and Krebs \cite{stephenskrebs1986foraging}. We characterise each patch type, $k$, with an expected utility return rate as a function of time spent within the patch, $g_k(t_k)$. We assume that patches are encountered randomly with rate $\lambda_k$ as Poisson processes. The forager's decision is now how long to spend in each patch type, with a strategy described as $\textbf{t} = [t_1, t_2, ... , t_k]$ ($t_i=0$ meaning the patch is ignored) . We can write the expected patch utility rate as

\begin{equation}
R_{media} = \dfrac{\sum_k \lambda_k g_k(t_k)}{1 + \sum_k \lambda_k t_k} \,.
\end{equation}

Similarly to the prey choice derivation, we differentiate with respect to the time spent in a patch type, $t_j$,

\begin{equation}
\frac{\partial R_{media}}{\partial t_j} = \dfrac{\lambda_j g'_j(t_j) (1 + \sum_k \lambda_k t_k) - \lambda_j (\sum_k \lambda_k g_k(t_k))}{(1 + \sum_k \lambda_k t_k)^2} \,,
\end{equation}

where $g'_j(t_j) = \frac{\partial g_j(t_j))}{\partial t_j}$. Setting this equal to zero, we find the maximum $R_{env}$ when 

\begin{equation}
    g'_j(t_j) = R_{env} \quad \quad \forall j \,. \label{eqn:charnov_criteria}
\end{equation}

This is Charnov's marginal value theorem \cite{Charnov1976Apr} and states that an optimal forager will leave a patch when the marginal utility rate of the patch equals the overall rate of utility from foraging in the environment. And foragers will not spend any time in a patch if the marginal rate never reaches the environmental rate i.e. $g'_j(t_j) < R_{env} \quad \forall t_j$. This makes sense intuitively --- time spent in a patch with rate $g_j$ carries an opportunity cost of time not spent foraging in the wider environment with utility rate $R_{env}$. 

We can find which patches will be visited using the "patches as prey" algorithm \cite{stephenskrebs1986foraging}. This is a similar algorithm to the diet choice model but with patches ranked in order of their maximum profitability, $\frac{g_k(t_k^*)}{t_k^*}$. patch types are added to the diet one at a time, with the marginal value theorem applied to all included patches after adding each new patch to recalculate the environmental utility rate. This is done with all patch types, or until Inequality \ref{eqn:charnov_criteria} fails. 

How would this model of patches effect the conclusions of the main paper? As in the main paper, we assume that media producers have an incentive to create information patches that attract and hold attention. People are still driven towards patches with high patch utility rates. If patch degradation occurs through consuming the most attractive items first then then there would still be a selective pressure toward high utility rate information items, as this would make the patch more attractive before degradation and keep foragers in the patch for longer as it degrades. And this pressure would still apply more strongly to short-form media than long-form media (due to more time switching between short-form media). The conclusions in the main paper would still follow, although the full model would be more complicated. We are confident that the conclusions would hold under any reasonable model of patch degradation. 

\section{Supplementary Information --- The Merged Poisson Process for Patches}

Here we justify using average values to describe the expected patch utility rates, instead of summations over information types. We have not seen this derivation before in the foraging literature, but it is relatively straightforward. The result is used without derivation in  \cite{pirolli2009information}. 

In the main text we write down an equation for the expected patch rate in terms of the characteristics of the information within the patch diet, $D$,

\begin{equation}
    R_{media} = \dfrac{\sum_{i \in D} \lambda_{i} u_{i}}{1 + \sum_{i \in D} \lambda_{i} t_{i}} \label{eqn:patch_rate_SI} \,.
\end{equation}

In this model, information types are encountered as independent Poisson processes with rates, $\lambda_i$, during time spent searching, with total searching time $T_s$. Items have utilities $u_i$ and handling times $t_i$. With some simple algebraic manipulation we can write down

\begin{equation}
    R_{media} = \dfrac{ (\sum_D \lambda_i) \frac{\sum_D \lambda_i u_i T_s}{\sum_D \lambda_i T_s}}{1 + (\sum_D \lambda_i) \frac{\sum_D \lambda_i t_i T_s}{\sum_D \lambda_i T_s}} \label{eqn:patch_rate_expanded} \,.
\end{equation} 

The rate of a combined Poisson process is equal to the sum of the rate of the independent Poisson processes, $\lambda_p = \sum_D \lambda_i$ \cite{gallager2012discrete}. 

We define the average utility of items encountered in the patch as the total utility gained divided by the total number of items handled, 

\begin{equation}
    \bar{u}_p = \frac{\sum_D \lambda_i u_i T_s}{\sum_D \lambda_i T_s} \,.    
\end{equation}

Similarly the average time spent handling items encountered is the total time spent handling divided by the number of items handled, 

\begin{equation}
    \bar{t}_p = \frac{\sum_D \lambda_i t_i T_s}{\sum_D \lambda_i T_s} \,.    
\end{equation}

Substituting these relations into equation \ref{eqn:patch_rate_expanded},

\begin{equation}
    R_{media} = \dfrac{\lambda_p \bar{u}_p}{1 + \lambda_p \bar{t}_p} \label{eqn:Holling_avg_SI}  \,.
\end{equation}

We can therefore replace the patch rate equation (equation \ref{eqn:patch_rate_SI}) with averages taken over the merged Poisson process. This is a variation of Holling's disc equation \cite{holling1959some}, considering average values. 

\section{Extended Data --- Full Statistical Results}

\subsection{Timeseries Analysis}

The Kwiatkowski–Phillips–Schmidt–Shin (KPSS) test considers a null hypothesis of no trend. This is a one-sided test. Table \ref{tbl:timeseries} reports the KPSS statistics and the p-values for each of the analysed categories in the Corpus of Historical American English (COHA). Exact p-values are difficult to calculate below 0.01 and are not provided by python's statsmodels package \cite{seabold2010statsmodels}, we have therefore denoted these as $<0.01$ where applicable.

The Mann-Kendall test is a non-parametric trend test with the null hypothesis of no trend. This is a two-sided test. We report (Table \ref{tbl:timeseries}) the normalised z-score, the p-value, Kendall's Tau, the Mann-Kendall score and slope. Exact p-values below 0.01 and are not provided by python's statsmodels package \cite{seabold2010statsmodels}, we have therefore denoted these as $<0.01$ where applicable.

\begin{table}[]
\begin{tabular}{|l|ll|}
\hline
            & \multicolumn{2}{l|}{Word Entropy}                                                                               \\ \hline
            & \multicolumn{1}{l|}{KPSS (KPSS Statistic, p-value)} & Mann-Kendall (z, p-value, Tau, MK score, slope)           \\ \hline
news        & \multicolumn{1}{l|}{(1.4725, \textbf{$<$0.01})}    & (7.5198, \textbf{$<$0.01}, 0.5157, 2451.0000, 0.0046)               \\ \hline
magazines   & \multicolumn{1}{l|}{(1.7361, \textbf{$<$0.01})}    & (10.9990, \textbf{$<$0.01},0.7172, 4144.0000, 0.0027)               \\ \hline
fiction     & \multicolumn{1}{l|}{(1.2372, \textbf{$<$0.01})}    & (7.5911,\textbf{$<$0.01}, 0.4927, 2900.0000, 0.0017)                \\ \hline
non-fiction & \multicolumn{1}{l|}{(1.4084, \textbf{$<$0.01})}    & (5.9100,\textbf{$<$0.01}, 0.3836, 2258.0000, 0.0019)                \\ \hline
            & \multicolumn{2}{l|}{Type Token Ratio}                                                                           \\ \hline
            & \multicolumn{1}{l|}{KPSS (KPSS Statistic, p-value)} & Mann-Kendall (z, p-value, Tau, MK score, slope)           \\ \hline
news        & \multicolumn{1}{l|}{(1.1982, \textbf{$<$0.01})}    & (5.3317, \textbf{$<$0.01},0.3657, 1738.0000, 0.0005)       \\ \hline
magazines   & \multicolumn{1}{l|}{(1.0223, \textbf{$<$0.01})}    & (5.9933, \textbf{$<$0.01}, 0.3908, 2258.0000, 0.0002)      \\ \hline
fiction     & \multicolumn{1}{l|}{(0.8972, \textbf{$<$0.01})}    & (5.9891, \textbf{$<$0.01}, 0.3887, 2288.0000, 0.0003)      \\ \hline
non-fiction & \multicolumn{1}{l|}{(0.6866, 0.0148)}               & (2.4774, 0.0132,, 0.1609, 947.0000, 0.0001)               \\ \hline
            & \multicolumn{2}{l|}{Zipf exponent}                                                                              \\ \hline
            & \multicolumn{1}{l|}{KPSS (KPSS Statistic, p-value)} & Mann-Kendall (z, p-value, Tau, MK score, slope)           \\ \hline
news        & \multicolumn{1}{l|}{(1.5085, \textbf{$<$0.01})}    & (-7.8083, \textbf{$<$0.01}, -0.5355, -2545.0000, -0.0002)  \\ \hline
magazines   & \multicolumn{1}{l|}{(1.7521, \textbf{$<$0.01})}    & (-11.4025, \textbf{$<$0.01}, -0.7435, -4296.0000, -0.0001) \\ \hline
fiction     & \multicolumn{1}{l|}{(1.3244, \textbf{$<$0.01})}    & (-7.5335, \textbf{$<$0.01}, -0.4890, -2878.0000, -0.0001)  \\ \hline
non-fiction & \multicolumn{1}{l|}{(1.2890, \textbf{$<$0.01})}    & (-6.1038, \textbf{$<$0.01}, -0.3962, -2332.0000, -0.0001)  \\ \hline
\end{tabular}
\caption{Timeseries analysis across different categories and measures for text samples from COHA between 1900 and 2009. In each cell, the p-value of a Kwiatkowski–Phillips–Schmidt–Shin (KPSS) test and a Mann Kendall (MK) test are shown respectively. Significant trends at $p<0.01$ are emboldened. For both tests, p-values below 0.01 mean we can reject the null hypothesis of stationarity at 1\% significance.}
    \label{tbl:timeseries}
\end{table}

\subsection{Differences in Media Categories}

We ran ANOVA tests to test for differences between media categories in each of the lexcical measures in the British National Corpus (BNC), Corpus of Contemporary American English (COCA), and the Corpous of Historical American English (restricted to 2000-2007). Results are shown in Table \ref{tbl:anova}.

\begin{table}[]
\begin{tabular}{|l|l|}
\hline
             & Word Entropy ANOVA       \\ \hline
COHA (DOF:3) & (F = 86, p = 7.68e-54)   \\ \hline
COCA (DOF:3) & (F = 37, p = 8.99e-22)   \\ \hline
BNC (DOF:2)  & (F = 689, p = 1.76e-205) \\ \hline
             & Type Token Ratio ANOVA   \\ \hline
COHA (DOF:3) & (F = 34, p = 5.95e-22)   \\ \hline
COCA (DOF:3) & (F = 19, p = 5.21e-12)   \\ \hline
BNC (DOF:2)  & (F = 425, p = 3.63e-143) \\ \hline
             & Zipf Exponent ANOVA      \\ \hline
COHA (DOF:3) & (F = 92, p = 2.14e-57)   \\ \hline
COCA (DOF:3) & (F = 41, p = 3.54e-24)   \\ \hline
BNC (DOF:2)  & (F = 712, p = 2.67e-210) \\ \hline
\end{tabular}
\caption{Analysis of differences in word measures across media categories within each text corpus. ANOVA tests are reported. All are significant.}
    \label{tbl:anova}
\end{table}

\section{Extended Data --- COHA Timeseries for Type Token Ratio and Zipf exponent}

\begin{figure}[ht]
		\centering
		\includegraphics[width=1\textwidth]{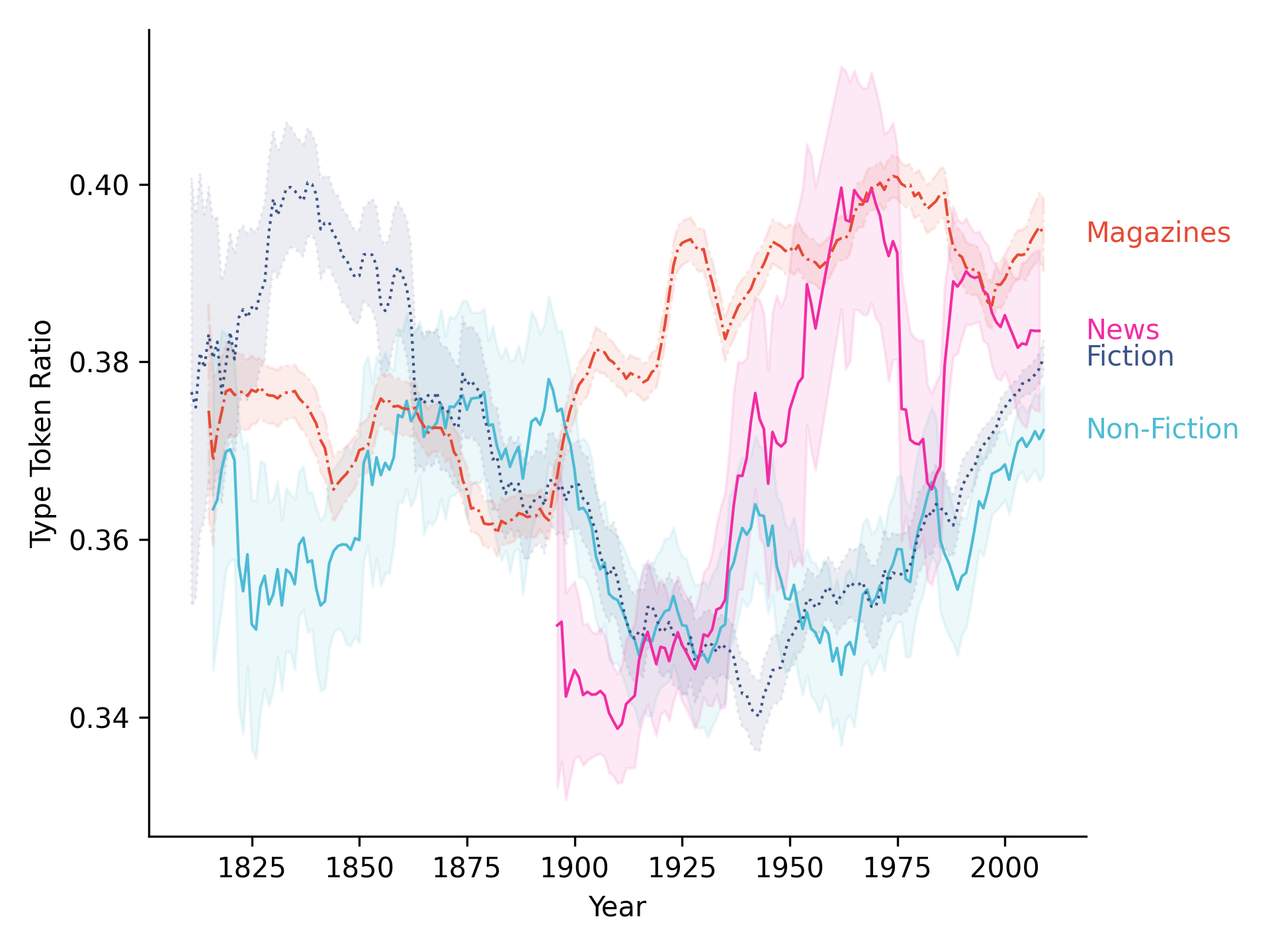}
		\caption{Historical timeseries of type token ratio in the Corpus of Historical American English. Type token ratio was calculated for text samples from COHA truncated with $N=2000$ words. For each media category and year, a moving average of all valid samples with $\pm 5$ years was calculated. The shaded region shows a 95\% confidence interval for this average.}
		\label{fig:timeseries-ttr}
	\end{figure}
	
\begin{figure}[ht]
		\centering
		\includegraphics[width=1\textwidth]{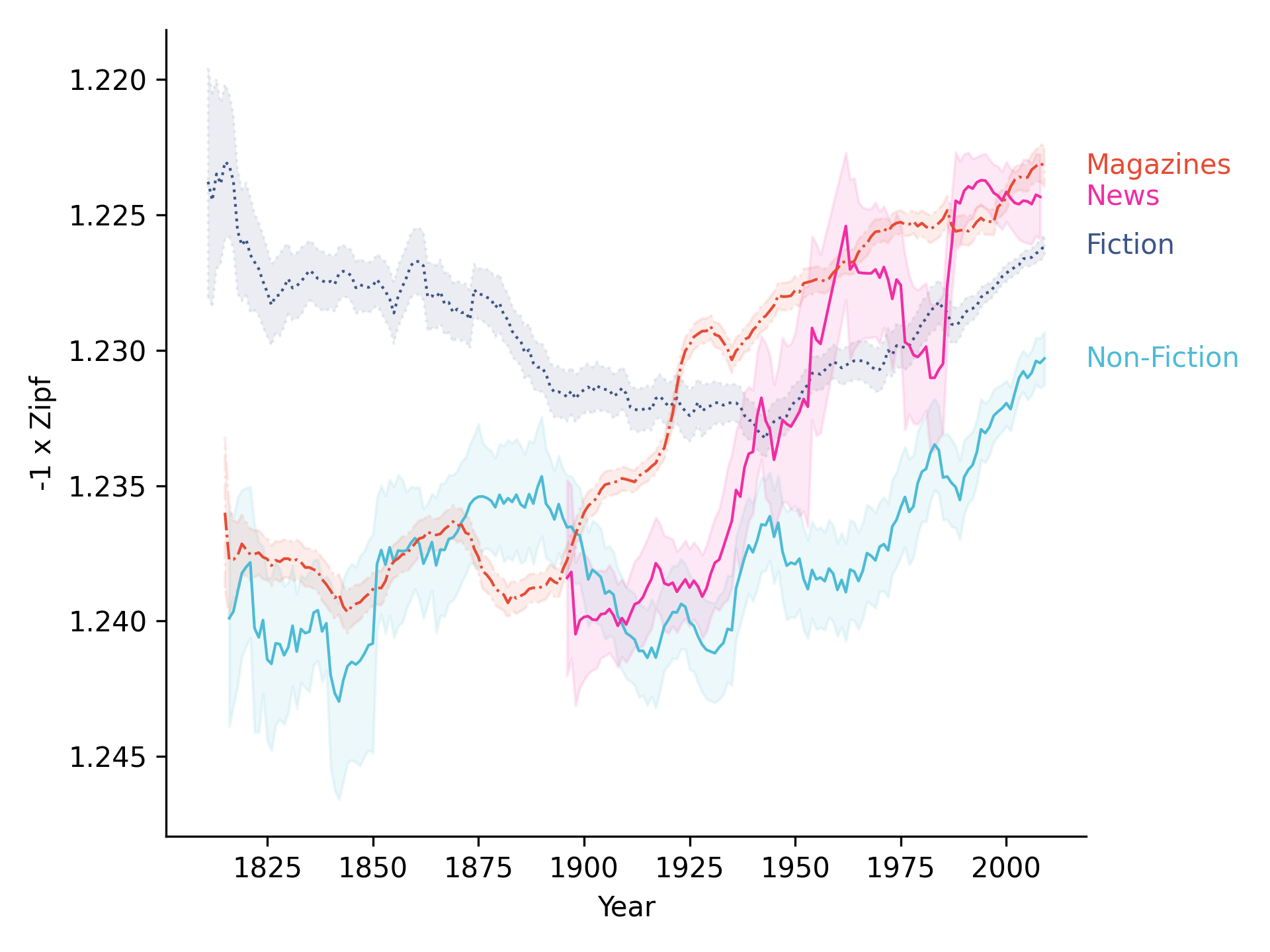}
		\caption{Historical timeseries of Zipf exponent in text samples in written media categories in American English. The timeseries was calculated in the same way as in the previous figure. }
		\label{fig:timeseries-zipf}
	\end{figure}

\section{Extended Data --- Corpora Boxplot Distributions for Word Entropy, Type Token Ratio and Zipf exponent}

\begin{figure}[ht]
		\centering
		\includegraphics[width=1\textwidth]{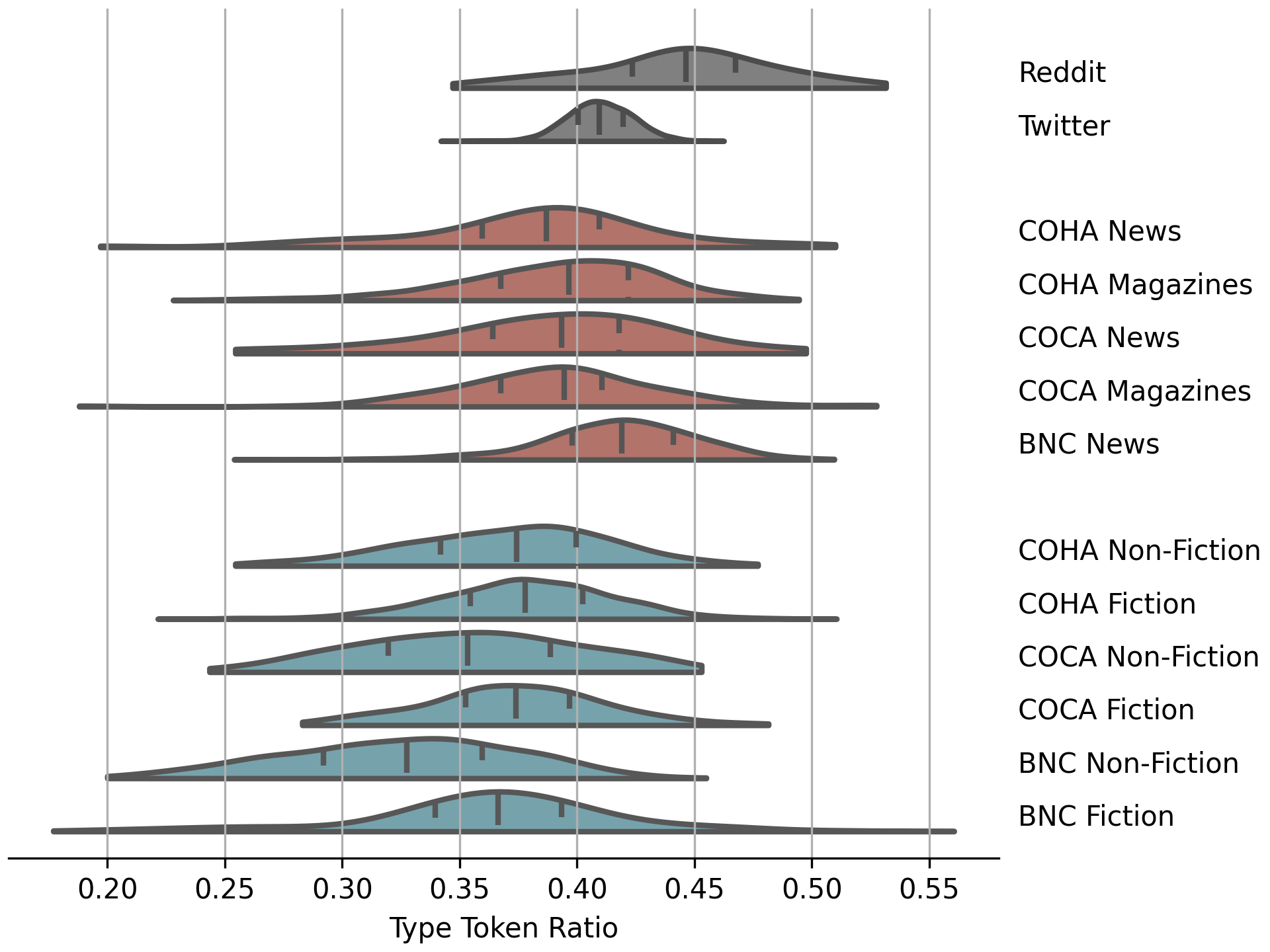}
		\caption{Distribution snapshots of type token ratio across different text corpora for text samples with  $N=2000$ words. COHA samples are from the year 2000 onwards only. Social media text samples were collated from status updates.}
		\label{fig:boxplots-ttr}
	\end{figure}
	
\begin{figure}[ht]
		\centering
		\includegraphics[width=1\textwidth]{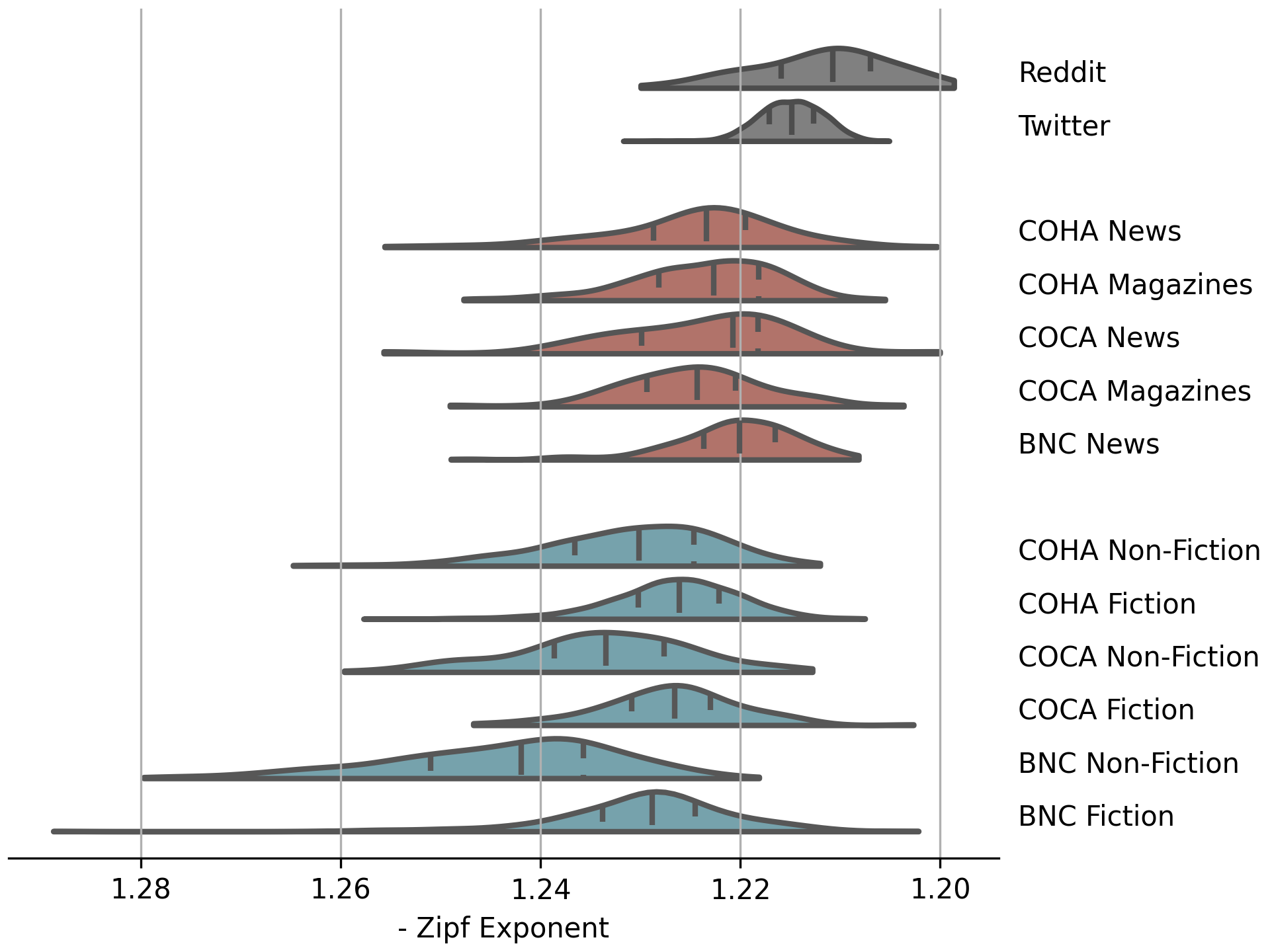}
		\caption{Distribution snapshots of the Zipf exponent across different text corpora for text samples with  $N=2000$ words. COHA samples are from the year 2000 onwards only. Social media text samples were collated from status updates.}
		\label{fig:boxplots-zipf}
	\end{figure}

\section{Supplementary - Timeseries Breakpoint Analysis}

\begin{figure}[ht]
		\centering
		\includegraphics[width=1\textwidth]{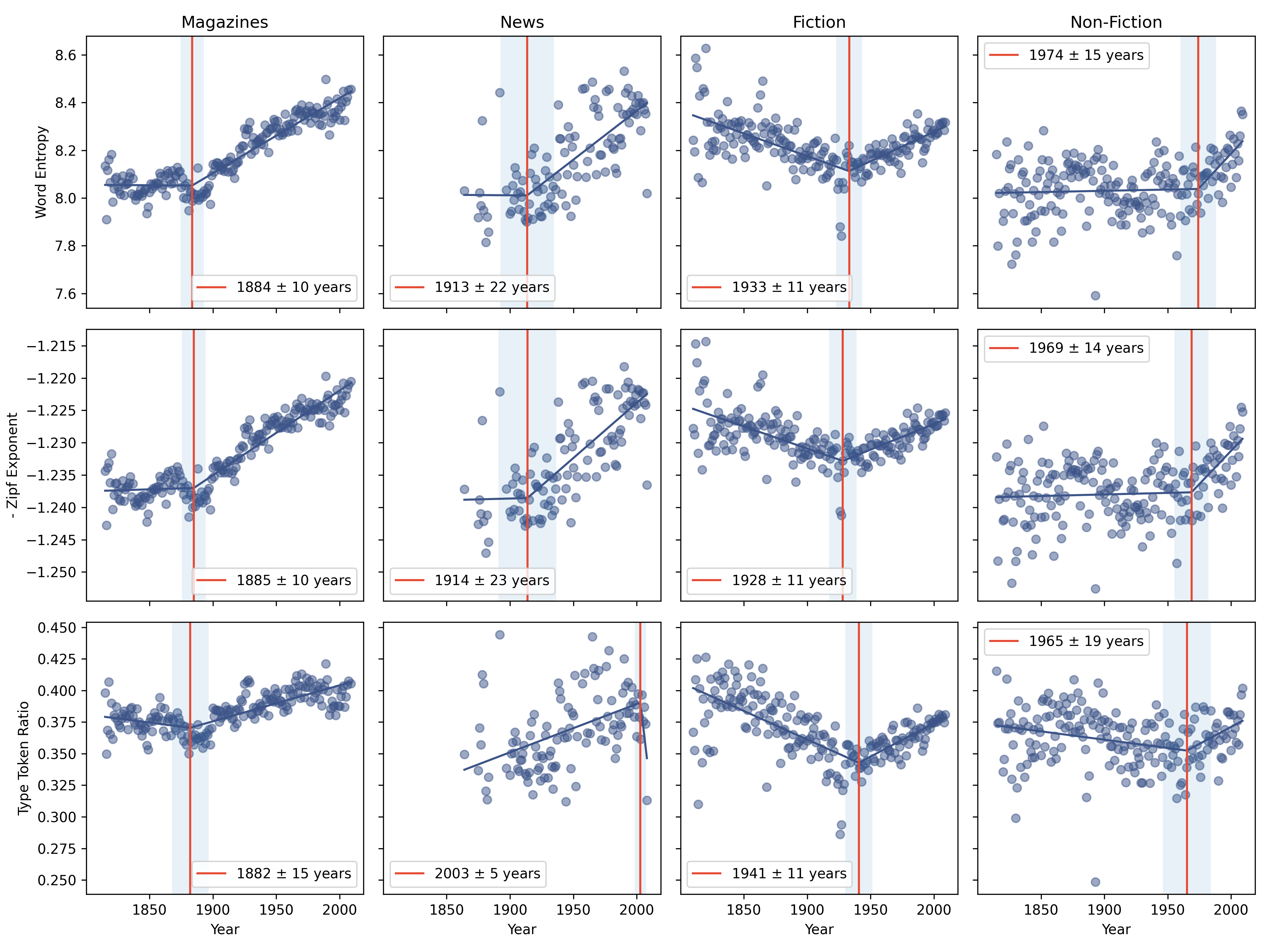}
		\caption{Median annual values for each category and lexical measure. The points were fit with a piecewise-regression, with red lines showing the estimated breakpoints. The shaded region shows a 95\% confidence interval for those breakpoints.}
		\label{fig:piecewise-all}
	\end{figure}

As discussed in Methods, we carried out a piecewise-regression analysis on the median annual values for each of the lexical measures and media categories (Figure \ref{fig:piecewise-all}). With the type token ration for the News media category, the breakpoint was found close to the edge of the data. If we restrict the position to avoid being close to the edge then the breakpoint is estimated in a similar location as to the Word Entropy and Zipf exponent. The short-form media shows signs of a rise in lexical diversity before long-form media, consistent with the model in the main paper. 

\begin{figure}[ht]
		\centering
		\includegraphics[width=1\textwidth]{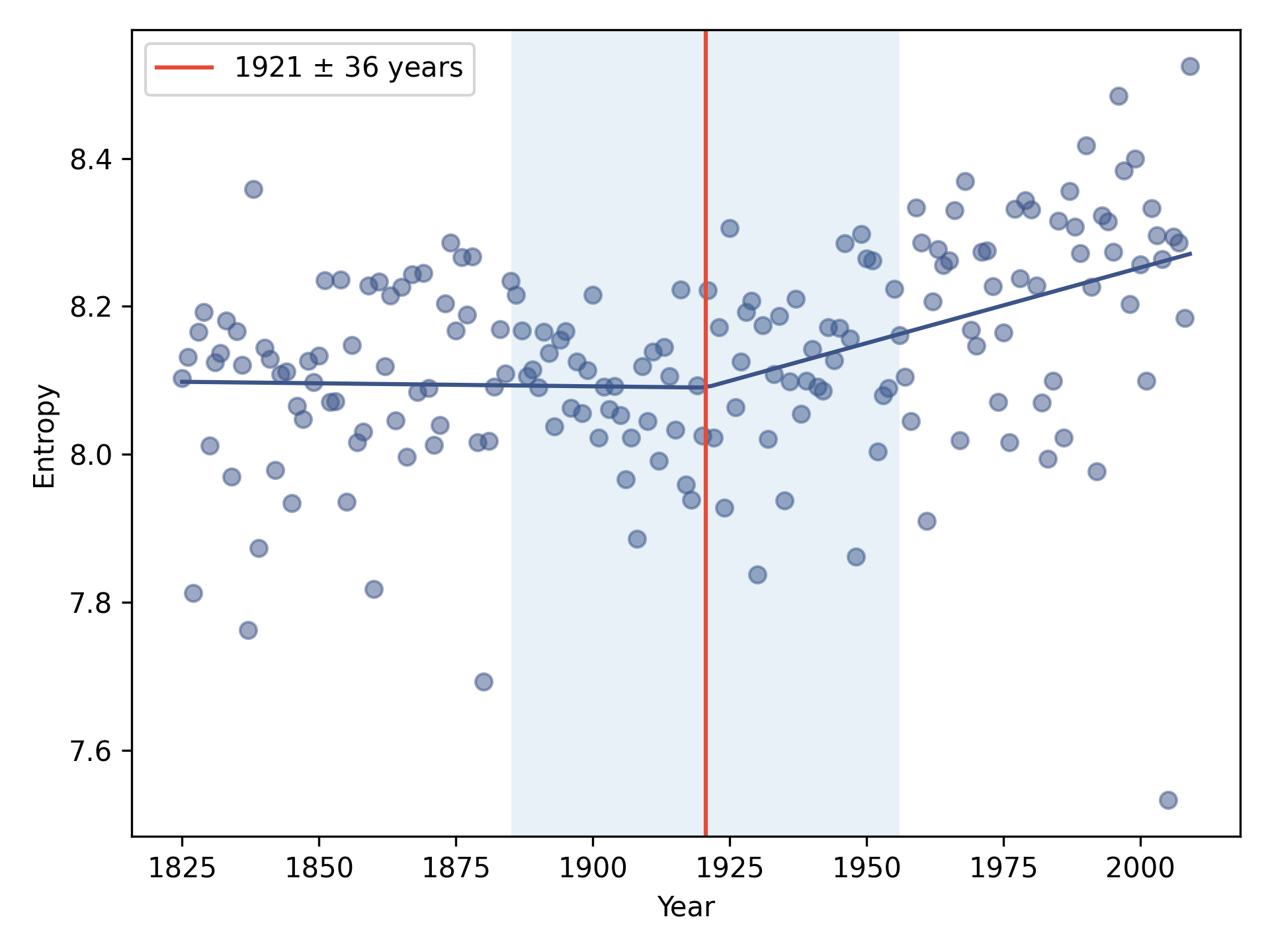}
		\caption{Mean annual values for the media categories combined for word entropy. Annual means were first found within each media category, and then averaged over the media categories. The points were fit with a piecewise-regression, with red lines showing the estimated breakpoint. The shaded region shows a 95\% confidence interval for that breakpoint.}
		\label{fig:piecewise-combined}
	\end{figure}

We ran the same analysis with the media categories collated to give an average mean each year (Figure \ref{fig:piecewise-combined}). Notably, the confidence interval for the breakpoint includes the year 1900. 

\end{document}